\documentclass[a4paper,12pt]{article}
\usepackage{epsfig}
\usepackage{color}
\usepackage[numbers,sort&compress]{natbib}

\newlength{\dinwidth}
\newlength{\dinmargin}
\newcommand{\spur}[1]{\not\! #1 \,}

\begin{document}
\title{Study of the exclusive $b\to u \ell^- \overline{\nu}_{\ell}$ decays  in the MSSM with and without R-parity violation}
\author{C. S. Kim\thanks{E-mail: cskim@yonsei.ac.kr}~ and
Ru-Min Wang\thanks{E-mail: ruminwang@cskim.yonsei.ac.kr}
%,~~Freddy Simanjuntak\thanks{E-mail: freddy@cskim.yonsei.ac.kr}
\\
{\it Department of Physics, Yonsei University, Seoul 120-479,
Korea}}

\maketitle \vspace{0.5cm}

\begin{abstract}

\noindent We study the exclusive $b \to u \ell^-
\overline{\nu}_{\ell}~(\ell=\tau,\mu,e)$ decays in the MSSM with and
without R-parity violation. {}From the experimental measurements of
branching ratios $\mathcal{B}(B^-_u\to \tau^-\overline{\nu}_\tau)$,
$\mathcal{B}(B^-_u\to M'^0\ell'^- \overline{\nu}_{\ell'})$ and
$\mathcal{B}(\overline{B}^0_d\to M'^+\ell'^-
\overline{\nu}_{\ell'})$ $(\ell'=\mu,e,~M'=\pi,\rho)$, we  derive
new upper bounds on the relevant new physics parameters within the
decays. Using the constrained  new physics  parameter spaces, we
predict the charged Higgs effects and the  R-parity violating
effects on the branching ratios, the normalized forward-backward
asymmetries of charged leptons, and the ratios of longitudinal to
transverse polarization of the vector mesons, which have not been
measured or have not been well measured yet. We find that the
charged Higgs effects and the R-parity violating effects could be
large and measurable in some cases. Our results could be used  to
probe new physics effects in the  leptonic decays as well as the
semileptonic decays, and will correlate with searches for direct
supersymmetric signals in future experiments.

\end{abstract}

\hspace{0.45cm}{PACS Numbers: 13.20.He  12.60.Jv 14.80.Cp 11.30.Fs}

\newpage
\section{Introduction}

The rare $B$ decays  have received a lot of attention, since they
are very promising for investigating the standard model (SM) and
searching for new physics (NP) beyond it. Among these $B$ decays,
the rare semileptonic ones have played a central role for a long
time, since the most precise measurements of the CKM matrix elements
$|V_{ub}|$ and $|V_{cb}|$ are based on the semileptonic decays $b\to
u \ell^-\overline{\nu}_\ell$ and $b\to c \ell^-\overline{\nu}_\ell$,
respectively. These decays can also be very useful to test the
various NP
 scenarios like the two Higgs doublet models \cite{2HDM}, the
minimal supersymmetric standard model (MSSM) \cite{MSSM1,MSSM2}, and
$etc$.

It is known that the charged Higgs boson exists in any models with
two or more Higgs doublets, such as the MSSM which contains two
Higgs doublets $H_u$ and $H_d$ coupling to up and down type quarks,
respectively.  The charged Higgs sectors of all these models may be
characterized by the ratio of the two Higgs vacuum expectation
values, $\tan\beta$, and the mass of the charged Higgs, $m_H$.
 Large $\tan\beta$ regime of
both supersymmetric and nonsupersymmetric models has a few
interesting signatures in $B$ physics (for instance, see  Refs.
\cite{largeTanbeta1,largeTanbeta2,largeTanbeta3,largeTanbeta4,largeTanbeta5,
largeTanbeta6,largeTanbeta7,BurasHcorrection,HiggsButotaunu} and references therein).
One of the most clear ones is the suppression of
$\mathcal{B}(B_u^-\to\tau^-\overline{\nu}_\tau)$ with respect to its
SM expectation \cite{HiggsButotaunu}. In the MSSM, the charged Higgs
contributions to the exclusive $b\to u \ell^- \overline{\nu}_{\ell}$
decays, including $B_u^-\to\tau^-\overline{\nu}_\tau$ decay, come
from  the $b$ quark transforms to a $u$ quark emitting a virtual
charged Higgs that manifests itself as a lepton-neutrino pair. In
this paper, we will present a correlated analysis of all these
exclusive  $b\to u \ell^- \overline{\nu}_{\ell}$ observables within
the large $\tan\beta$ limit of the MSSM.

In the MSSM, one can introduce a discrete symmetry, called
$R$-parity ($R_p$) \cite{R-parity}, to enforce in a simple way the
lepton number ($L$) and the baryon number ($B$) conservations.  In view
of the important phenomenological differences between supersymmetric
models with and without $R_p$ violation, it is also worth studying
the extent to which $R_p$ can be broken. The effects of  SUSY
with $R_p$ violation in $B$ meson decays have been extensively
investigated, for instance Refs.
\cite{RPVstudy1,RPVstudy2,RPVstudy3,RPVstudy4,RPVstudy5,RPVstudy6,RPVLFV}.
In Ref. \cite{RPVLFV}, the $R_p$ violating (RPV) and lepton flavor
violating coupling effects have been studied in $B^-\to
\ell^-\overline{\nu}_\ell$ decays. The exclusive $b\to u \ell^-
\overline{\nu}_{\ell}$ decays involve the same set of the RPV coupling
products for every generation of leptons. In this work, still
assuming lepton flavor conservation, we will investigate the
sensitivity of the exclusive $b\to u \ell^- \overline{\nu}_{\ell}$
decays to the RPV coupling contributions  in the RPV MSSM, too.

The paper is organized as follows. In section 2,  we introduce the
theoretical frame of the exclusive $b\to u \ell^-
\overline{\nu}_{\ell}$ decays in the MSSM with and without $R_p$
violation in detail. In section 3, we tabulate all the theoretical
inputs. In sections 4 and 5,
 we deal with the numerical results. We display the constrained
parameter spaces which satisfy all the available experimental data,
and then we use the constrained parameter spaces to  predict the NP
effects on other quantities, which have not been measured or have
not been well measured yet. Section 6 contains our summary and
conclusion.

\section{The exclusive $b\to u \ell^- \overline{\nu}_{\ell}$ decays in the MSSM with and without R-parity violation}

In supersymmetric extensions of the SM, there are gauge invariant
interactions which violate the $B$ and the $L$ in general. To
prevent occurrences of these $B$ and $L$ violating interactions in
supersymmetric extensions of the SM, the additional global symmetry
is required. This requirement leads to the consideration of the so
called $R_p$ conservation (RPC).

\begin{figure}[hb]
\begin{center}
\includegraphics[scale=0.8]{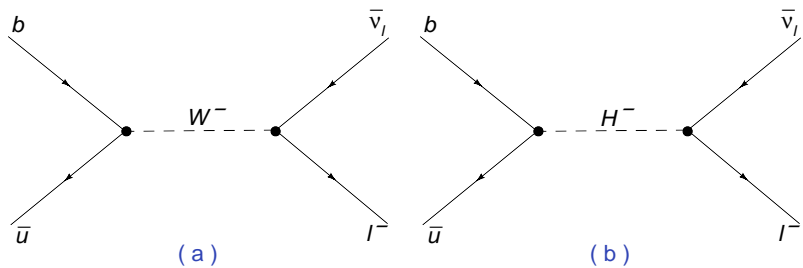}
\end{center}\vspace{-1cm}
\caption{The decays  $b\to u\ell^-\overline{\nu}_{\ell}$ are mediated
by a $W$ boson exchange in the SM, and in extensions of the SM also
by a charged Higgs exchange.}\label{WHexchange}
\end{figure}

In the MSSM with RPC, the terms in the effective Hamiltonian
relevant for the $b \to u \ell^- \overline{\nu}_{\ell}$ decays are
\begin{eqnarray}
\mathcal{H}^{R_p}_{eff}(b \to u\ell^-
\overline{\nu}_{\ell})=\frac{G_F}{\sqrt{2}}V_{ub}[(\overline{u}\gamma_\mu(1-\gamma_5)b)
(\overline{\ell}\gamma^\mu(1-\gamma_5)\nu_\ell)
-R_l(\overline{u}(1+\gamma_5)b)
 (\overline{\ell}(1-\gamma_5)\nu_{\ell})], \label{HTRP}
\end{eqnarray}
here
$R_l=\frac{\tan^2\beta}{m^2_{H}}\frac{\overline{m}_bm_l}{1+\epsilon_0\tan\beta}$,
parameter $\epsilon_0$ is generated at the one loop level (with the
main contribution originating from gluino diagrams). Note that
$\tilde{\epsilon}_0$ of \cite{BurasHcorrection} corresponds to
$\epsilon_0$ in our convention. The first term in Eq. (\ref{HTRP})
gives the SM contribution shown in Fig. \ref{WHexchange}(a), and the
second one gives that of the charged Higgs scalars shown in Fig.
\ref{WHexchange}(b).

Even though the requirement of RPC makes a theory consistent with
present experimental searches, there is no good theoretical
justification for this requirement. Therefore, the most general
models with explicit $R_p$ violation should be also considered. In
the most general superpotential of the MSSM, the RPV superpotential
is given by \cite{RPVSW}
\begin{eqnarray}
\mathcal{W}_{\spur{R_p}}&=&\mu_i\hat{L}_i\hat{H}_u+\frac{1}{2}
\lambda_{[ij]k}\hat{L}_i\hat{L}_j\hat{E}^c_k+
\lambda'_{ijk}\hat{L}_i\hat{Q}_j\hat{D}^c_k+\frac{1}{2}
\lambda''_{i[jk]}\hat{U}^c_i\hat{D}^c_j\hat{D}^c_k, \label{rpv}
\end{eqnarray}
where $\hat{L}$ and $\hat{Q}$ are the SU(2) doublet lepton and quark
superfields, respectively, $\hat{E}^c$, $\hat{U}^c$ and $\hat{D}^c$
are the singlet superfields, while $i$, $j$ and $k$ are generation
indices and the superscript $c$ denotes a charge conjugate field.

\begin{figure}[b]
\begin{center}
\includegraphics[scale=0.8]{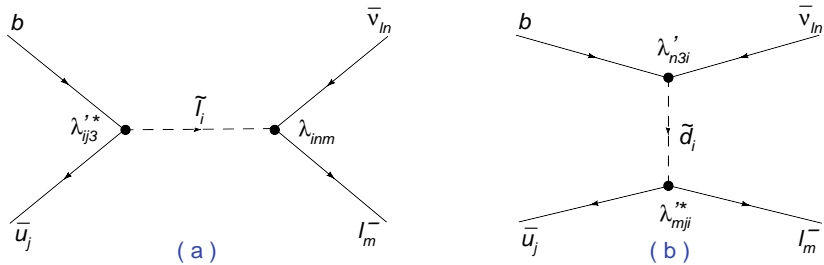}
\end{center}\vspace{-1cm}
\caption{ The RPV contributions to the exclusive $b \to u_j \ell^-_m
\overline{\nu}_{\ell n}$ decays  due to sleptons and squarks
exchange.}\label{RPVfig}
\end{figure}

{}From Eq. (\ref{rpv}), we can obtain the relevant four fermion
effective Hamiltonian for the $b \to u_j \ell^-_m
\overline{\nu}_{\ell n}$ processes with RPV couplings due to the
squarks and sleptons exchange
\begin{eqnarray}
 \mathcal{H}_{eff}(b \to u_j \ell^-_m \overline{\nu}_{\ell
n})^{\spur{R_p}}&=&-\sum_i \frac{\lambda'_{n3i}\lambda'^{*}_{mji}}
 {8m^2_{\tilde{d}_{iR}}}(\overline{u}_j\gamma_\mu(1-\gamma_5)b)(\overline{\ell}_m\gamma^\mu(1-\gamma_5){\nu}_{\ell n}) \nonumber\\
 && +\sum_i\frac{\lambda_{inm}\lambda'^{*}_{ij3}}{4m^2_{\tilde{\ell}_{iL}}}(\overline{u}_j(1+\gamma_5)b)
 (\overline{\ell}_m(1-\gamma_5)\nu_{\ell n}).\label{RPVHeff}
\end{eqnarray}
The corresponding RPV feynman diagrams for the $b \to u_j \ell^-_m
\overline{\nu}_{\ell n}$ processes  are displayed in Fig.
\ref{RPVfig}. Note that the operators in Eq. (\ref{RPVHeff}) take
the same form as those of the MSSM with RPC shown in Eq.
(\ref{HTRP}).

Then, we can obtain the total effective Hamiltonian for the $b\to u
\ell^- \overline{\nu}_{\ell}$ processes in the RPV MSSM
\begin{eqnarray}
 \mathcal{H}^{\spur{R_p}}_{eff}(b\to u \ell^- \overline{\nu}_{\ell})&\equiv&
 \mathcal{H}_{eff}(b\to u \ell^- \overline{\nu}_{\ell})^{\rm SM}+
 \mathcal{H}_{eff}(b\to u \ell^- \overline{\nu}_{\ell})^{\spur{R_p}} \nonumber\\
&=&\left(\frac{G_F}{\sqrt{2}}V_{ub}-\sum_i
\frac{\lambda'_{n3i}\lambda'^{*}_{mji}}
 {8m^2_{\tilde{d}_{iR}}}\right)(\overline{u}_j\gamma_\mu(1-\gamma_5)b)(\overline{\ell}_m\gamma^\mu(1-\gamma_5){\nu}_{\ell n}) \nonumber\\
 && +\sum_i\frac{\lambda_{inm}\lambda'^{*}_{ij3}}{4m^2_{\tilde{\ell}_{iL}}}~(\overline{u}_j(1+\gamma_5)b)
 (\overline{\ell}_m(1-\gamma_5)\nu_{\ell n}).\label{HTRPV}
\end{eqnarray}
Based on the effective Hamiltonian in Eq.
(\ref{HTRPV}), we will give the expressions of physical quantities for the RPV MSSM
later in detail. Note that the operators in Eq. (\ref{HTRPV}) have the exactly
same form as those of the MSSM with RPC shown in Eq. (\ref{HTRP}). For
the expressions of the charged Higgs contributions, we just need let
$\sum_i \frac{\lambda'_{n3i}\lambda'^{*}_{mji}}
 {8m^2_{\tilde{d}_{iR}}}=0$ and replace
 $\sum_i\frac{\lambda_{inm}\lambda'^{*}_{ij3}}{4m^2_{\tilde{\ell}_{iL}}}$ with $-\frac{G_F}{\sqrt{2}}V_{ub}R_l$.
In the following expressions and numerical analysis, we will keep
the masses of all three generation charged leptons, but ignore all
neutrino masses.

\subsection{The branching ratio for $B^-_u \to \ell^- \overline{\nu}_\ell$}

$B^-_u \to \ell^- \overline{\nu}_{\ell}$ decay amplitude can be
obtained in terms of Eq. (\ref{HTRPV}),
\begin{eqnarray}
\mathcal{M}^{\spur{R_p}}(B^-_u \to \ell^- \overline{\nu}_{\ell})
&=&\langle \ell^- \overline{\nu}_\ell|\mathcal{H}^{\spur{R_p}}_{eff}(b\to u\ell^-\overline{\nu}_{\ell})|B^-\rangle \nonumber\\
&=&\left[\frac{G_F}{\sqrt{2}}V_{ub}-\sum_i
\frac{\lambda'_{n3i}\lambda'^{*}_{m1i}}
 {8m^2_{\tilde{d}_{iR}}}\right]\langle 0|\bar{u}\gamma_\mu(1-\gamma_5)b|B^-\rangle \overline{\ell}_{m}\gamma^\mu(1-\gamma_5)\nu_{\ell n}\nonumber\\
&&+\sum_i\frac{\lambda_{inm}\lambda'^{*}_{i13}}{4m^2_{\tilde{\ell}_{iL}}}\langle
0|\bar{u}(1+\gamma_5)b|B^-\rangle
\overline{\ell}_{m}(1-\gamma_5)\nu_{\ell n}.
\end{eqnarray}
After using the definitions of $B$ meson decay constant
\cite{BallZwicky}
\begin{eqnarray}
&&\langle 0|\bar{u}\gamma_\mu\gamma_5b|B^-\rangle =if_{_{B_u}}p_{_{B}\mu},\\
{\rm and}~&&\langle 0|\bar{u}\gamma_5b|B^-\rangle
=-if_{_{B_u}}\mu_{_{B_u}}~~~~\mbox{with}~~~~\mu_{_{B_u}}
\equiv\frac{m_{B_u}^2}{\overline{m}_b+\overline{m}_u},
\end{eqnarray}
we get the branching ratio for $B^-_u \to \ell^-
\overline{\nu}_{\ell}$
\begin{eqnarray}
\mathcal{B}^{\spur{R_p}}(B^-_u \to \ell^-
\overline{\nu}_{\ell})=\left|\frac{G_F}{\sqrt{2}}V_{ub}-\sum_i
\frac{\lambda'_{n3i}\lambda'^{*}_{m1i}}
 {8m^2_{\tilde{d}_{iR}}}+\sum_i\frac{\lambda_{inm}\lambda'^{*}_{i13}}{4m^2_{\tilde{\ell}_{iL}}}\frac{\mu_{_{B_u}}}{m_\ell}\right|^2
 \frac{\tau_{_{B_u}}}{4\pi}f_{_{B_u}}^2m_{_{B_u}}m^2_\ell\left[1-\frac{m_\ell^2}{m^2_{B_u}}\right]^2.
\end{eqnarray}
{}From the above expression,  we note that, unlike the contributions
of squark exchange coupling
$\lambda'_{n3i}\lambda'^{*}_{m1i}$  and  the SM to $\mathcal{B}(B^-_u \to
\ell^-\overline{\nu}_{\ell})$, slepton exchange coupling
$\lambda_{inm}\lambda'^{*}_{i13}$ is not suppressed by $m^2_{\ell}$.

\subsection{The branching ratio for $B_{(s)} \to P \ell^- \overline{\nu}_\ell~~~(P=\pi,K)$}

$B \to P\ell^- \overline{\nu}_{\ell}$ decay amplitude can be
written as
\begin{eqnarray}
\mathcal{M}^{\spur{R_p}}(B \to P \ell^- \overline{\nu}_{\ell})
&=&\langle P\ell^- \overline{\nu}_\ell|\mathcal{H}^{\spur{R_p}}_{eff}(b\to u\ell^-\overline{\nu}_{\ell})|B\rangle\nonumber\\
&=&\left[\frac{G_F}{\sqrt{2}}V_{ub}-\sum_i
\frac{\lambda'_{n3i}\lambda'^{*}_{m1i}}
 {8m^2_{\tilde{d}_{iR}}}\right]\langle P|\bar{u}\gamma_\mu(1-\gamma_5)b|B\rangle \overline{\ell}_{m}\gamma^\mu(1-\gamma_5)\nu_{\ell n}\nonumber\\
&&+\sum_i\frac{\lambda_{inm}\lambda'^{*}_{i13}}{4m^2_{\tilde{\ell}_{iL}}}\langle
P|\bar{u}(1+\gamma_5)b|B\rangle
\overline{\ell}_{m}(1-\gamma_5)\nu_{\ell n}.
\end{eqnarray}
Using the $B \to P$ transition form factors
\cite{BallZwicky}
\begin{eqnarray}
&&c_P\langle P(p)|\bar{u}\gamma_{\mu}b|B (p_{_{B}})\rangle
=f^P_+(s)(p+p_{_{B}})_\mu+\left[f^P_0(s)-f^P_+(s)\right]\frac{m^2_B-m^2_P}{s}q_\mu,\\
&&c_P\langle P(p)|\bar{u}b|B
(p_{_{B}})\rangle=f^P_0(s)\frac{m^2_B-m^2_P}{\overline{m}_b-\overline{m}_u},
\end{eqnarray}
where the factor $c_P$ accounts for the flavor content of particles
($c_P=\sqrt{2}$ for $\pi^0$, and $c_P=1$ for $\pi^-,K^-$) and
$s=q^2~~(q=p_{_{B}}-p)$, the differential branching ratio for $B \to
P \ell^- \overline{\nu}_\ell$ is
\begin{eqnarray}
\frac{d\mathcal{B}^{\spur{R_p}}(B \to P
\ell^-\overline{\nu}_{\ell})}{ds~dcos\theta}&=&\frac{\tau_{_{B}}\sqrt{\lambda_P}}{2^7\pi^3m^3_Bc^2_P}\left(1-\frac{m_\ell^2}{s}\right)^2
\left[N_0^P+N_1^Pcos\theta+N_2^Pcos^2\theta\right],\hspace{3cm}
\label{brBtoPlnu}
\end{eqnarray}
{\small
\begin{eqnarray}
N_0^P&=&\left|\frac{G_F}{\sqrt{2}}V_{ub}-\sum_i
\frac{\lambda'_{n3i}\lambda'^{*}_{m1i}}
 {8m^2_{\tilde{d}_{iR}}}\right|^2\left[f^P_+(s)\right]^2\lambda_P\nonumber\\
 &&\hspace{-0.4cm}+\left|\frac{G_F}{\sqrt{2}}V_{ub}-\sum_i
\frac{\lambda'_{n3i}\lambda'^{*}_{m1i}}
 {8m^2_{\tilde{d}_{iR}}}+\sum_i\frac{\lambda_{inm}\lambda'^{*}_{i13}}{4m^2_{\tilde{\ell}_{iL}}}
 \frac{s}{m_\ell(\overline{m}_b-\overline{m}_u)}\right|^2m_\ell^2
\left[f^P_0(s)\right]^2\frac{(m_B^2-m^2_P)^2}{s},\label{NPcos0}\\
N_1^P&=&\left\{\left|\frac{G_F}{\sqrt{2}}V_{ub}-\sum_i
\frac{\lambda'_{n3i}\lambda'^{*}_{m1i}}
 {8m^2_{\tilde{d}_{iR}}}\right|^2+Re\left[\left(\frac{G_F}{\sqrt{2}}V_{ub}-\sum_i
\frac{\lambda'_{n3i}\lambda'^{*}_{m1i}}
 {8m^2_{\tilde{d}_{iR}}}\right)^\dagger\sum_i\frac{\lambda_{inm}\lambda'^{*}_{i13}}{4m^2_{\tilde{\ell}_{iL}}}
 \frac{s}{m_\ell(\overline{m}_b-\overline{m}_u)} \right] \right\}\nonumber\\
 &&\times~2m_\ell^2f^P_0(s)f^P_+(s)\sqrt{\lambda_P}\frac{(m_B^2-m^2_P)}{s},\label{NPcos1}\\
N_2^P&=&-\left|\frac{G_F}{\sqrt{2}}V_{ub}-\sum_i
\frac{\lambda'_{n3i}\lambda'^{*}_{m1i}}
 {8m^2_{\tilde{d}_{iR}}}\right|^2\left[f^P_+(s)\right]^2\lambda_P\left(1-\frac{m_\ell^2}{s}\right),\label{NPcos2}
\end{eqnarray}}
where $\theta$ is the angle between the momentum of $B$ meson and
the charged lepton in the c.m. system of $\ell$-$\nu$, and the
kinematic factor
$\lambda_P=m_B^4+m^4_P+s^2-2m_B^2m_P^2-2m_B^2s-2m_P^2s$.

Here, we give the definition of the normalized forward-backward (FB)
asymmetry of charged lepton \cite{NAFB}, which is more useful from
the experimental point of view,
\begin{eqnarray}
\overline{\mathcal{A}}_{FB}=\frac{\int^{+1}_{0}\frac{d^2\mathcal{B}}{dsdcos\theta}dcos\theta
-\int^{0}_{-1}\frac{d^2\mathcal{B}}{dsdcos\theta}dcos\theta}
{\int^{+1}_{0}\frac{d^2\mathcal{B}}{dsdcos\theta}dcos\theta
+\int^{0}_{-1}\frac{d^2\mathcal{B}}{dsdcos\theta}dcos\theta}.\label{NAFB}
\end{eqnarray}
Explicitly, for $B \to P \ell^- \overline{\nu}_\ell$ the normalized
FB asymmetry is
\begin{eqnarray}
\overline{\mathcal{A}}_{FB}(B \to P \ell^-
\overline{\nu}_\ell)=\frac{N^P_1}{2N^P_0+2/3N^P_2}~.\label{NAFBP}
\end{eqnarray}

\subsection{The branching ratio for $B_{(s)} \to V \ell^- \overline{\nu}_\ell~~~(V=\rho,K^{*})$}

Similarly, the expression for $B \to V\ell^- \overline{\nu}_{\ell}$
decay amplitude is
\begin{eqnarray}
\mathcal{M}^{\spur{R_p}}(B^-_u \to V \ell^- \overline{\nu}_{\ell})
&=&\langle V\ell^- \overline{\nu}_\ell|\mathcal{H}^{\spur{R_p}}_{eff}(b\to u\ell^-\overline{\nu}_{\ell})|B^-\rangle \nonumber\\
&=&\left[\frac{G_F}{\sqrt{2}}V_{ub}-\sum_i
\frac{\lambda'_{n3i}\lambda'^{*}_{m1i}}
 {8m^2_{\tilde{d}_{iR}}}\right]\langle V|\bar{u}\gamma_\mu(1-\gamma_5)b|B^-\rangle \overline{\ell}_{m}\gamma^\mu(1-\gamma_5)\nu_{\ell n}\nonumber\\
&&+\sum_i\frac{\lambda_{inm}\lambda'^{*}_{i13}}{4m^2_{\tilde{\ell}_{iL}}}\langle
V|\bar{u}(1+\gamma_5)b|B^-\rangle
\overline{\ell}_{m}(1-\gamma_5)\nu_{\ell n}.
\end{eqnarray}
In terms of the $B \to V$ form factors \cite{BallZwicky}
\begin{eqnarray}
c_V\langle
V(p,\varepsilon^{\ast})|\bar{u}\gamma_{\mu}(1-\gamma_5)b|B
(p_{_{B}})\rangle &=& \frac{2V^{V}(s)}{m_B+m_V}
\epsilon_{\mu\nu\alpha\beta}\varepsilon^{\ast\nu}p_{_B}^{\alpha}p^{\beta}\nonumber\\
&&-i\left[\varepsilon_{\mu}^\ast(m_B+m_V)A_1^{V}(s)
-(p_{_B}+p)_{\mu}({\varepsilon^\ast}\cdot{p_{_B}})\frac{A_2^{V}(s)}
{m_B+m_V}\right]\nonumber \\
&&+iq_{\mu}({\varepsilon^\ast}\cdot{p_{_B}})\frac{2m_V}{s}
[A_3^{V}(s)-A_0^{V}(s)],\\
 c_V\langle
V(p,\varepsilon^{\ast})|\bar{u}\gamma_5b|B (p_{_B})\rangle
&=&-i\frac{{\varepsilon^\ast}\cdot
p_{_B}}{m_B}\frac{2m_Bm_V}{\overline{m}_b+\overline{m}_u}A_0^V(s),
\end{eqnarray}
where $c_V=\sqrt{2}$ for $\rho^0$, $c_V=1$ for $\rho^-,K^{*-}$  and with the relation
$A_3^{V}(s)=\frac{m_B+m_V}{2m_V}A_1^{V}(s)-\frac{m_B-m_V}{2m_V}A_2^{V}(s),$
we have
\begin{eqnarray}
\frac{d\mathcal{B}^{\spur{R_p}}(B \to V
\ell^-\overline{\nu}_{\ell})}{ds~dcos\theta}&=&\frac{\tau_{_B}\sqrt{\lambda_V}}{2^7\pi^3m^3_Bc^2_V}\left(1-\frac{m_\ell^2}{s}\right)^2
\left[N_0^V+N_1^Vcos\theta+N_2^Vcos^2\theta\right],\hspace{3cm}
\end{eqnarray}
{\small
\begin{eqnarray}
N_0^V&=&\left|\frac{G_F}{\sqrt{2}}V_{ub}-\sum_i
\frac{\lambda'_{n3i}\lambda'^{*}_{m1i}}
 {8m^2_{\tilde{d}_{iR}}}\right|^2
 \Bigg\{\left[A^V_1(s)\right]^2\left(\frac{\lambda_V}{4m^2_V}+(m^2_\ell+2s)\right)(m_B+m_V)^2\nonumber\\
&&+\left[A^V_2(s)\right]^2\frac{\lambda_V^2}{4m_V^2(m_B+m_V)^2}
+\left[V^V(s)\right]^2\frac{\lambda_V}{(m_B+m_V)^2}(m^2_\ell+s)\nonumber\\
&&-A^V_1(s)A^V_2(s)\frac{\lambda_V}{2m^2_V}(m^2_B-s-m_V^2)\Bigg\}\nonumber\\
 &&+\left|\frac{G_F}{\sqrt{2}}V_{ub}-\sum_i
\frac{\lambda'_{n3i}\lambda'^{*}_{m1i}}
 {8m^2_{\tilde{d}_{iR}}}+\sum_i\frac{\lambda_{inm}\lambda'^{*}_{i13}}{4m^2_{\tilde{\ell}_{iL}}}
 \frac{s}{m_\ell(\overline{m}_b+\overline{m}_u)}\right|^2
 \left[A^V_0(s)\right]^2\frac{m_\ell^2}{s}\lambda_V,\label{NVcos0}\\
 N^V_1&=&\left\{\left|\frac{G_F}{\sqrt{2}}V_{ub}-\sum_i
\frac{\lambda'_{n3i}\lambda'^{*}_{m1i}}
 {8m^2_{\tilde{d}_{iR}}}\right|^2+Re\left[\left(\frac{G_F}{\sqrt{2}}V_{ub}-\sum_i
\frac{\lambda'_{n3i}\lambda'^{*}_{m1i}}
 {8m^2_{\tilde{d}_{iR}}} \right)^\dag\sum_i\frac{\lambda_{inm}\lambda'^{*}_{i13}}{4m^2_{\tilde{\ell}_{iL}}}
 \frac{s}{m_\ell(\overline{m}_b+\overline{m}_u)}
 \right]\right\}\nonumber\\
&&\times\left[
A^V_0(s)A^V_1(s)\frac{m^2_\ell(m_B+m_V)(m^2_B-m^2_V-s)\sqrt{\lambda_V}}{sm_V}
-A^V_0(s)A^V_2(s)\frac{m_\ell^2\lambda_V^{\frac{3}{2}}}{sm_V(m_B+m_V)}\right]\nonumber\\
&&+\left|\frac{G_F}{\sqrt{2}}V_{ub}-\sum_i
\frac{\lambda'_{n3i}\lambda'^{*}_{m1i}}
 {8m^2_{\tilde{d}_{iR}}}\right|^2A^V_1(s)V^V(s)~4s\sqrt{\lambda_V},\label{NVcos1}\\
N^V_2&=&-\left|\frac{G_F}{\sqrt{2}}V_{ub}-\sum_i
\frac{\lambda'_{n3i}\lambda'^{*}_{m1i}}
 {8m^2_{\tilde{d}_{iR}}}\right|^2\left(1-\frac{m^2_\ell}{s}\right)\lambda_V
 \left\{\left[A^V_1(s)\right]^2\frac{(m_B+m_V)^2}{4m_V^2}\right.\nonumber\\
&&\left.+\left[V^V(s)\right]^2\frac{s}{(m_B+m_V)^2}+\left[A^V_2(s)\right]^2\frac{\lambda_V}{4m_V^2(m_B+m_V)^2}
-A^V_1(s)A^V_2(s)\frac{m_B^2-m_V^2-s}{2m_V^2}\right\},
 \end{eqnarray}}
where $\lambda_V=m_B^4+m^4_V+s^2-2m_B^2m_V^2-2m_B^2s-2m_V^2s$.

{}From Eq. (\ref{NAFB}), the normalized FB asymmetry of $B \to V
\ell^- \overline{\nu}_\ell$ can be written as
\begin{eqnarray}
\overline{\mathcal{A}}_{FB}(B \to V \ell^-
\overline{\nu}_\ell)=\frac{N^V_1}{2N^V_0+2/3N^V_2}~.\label{NAFBV}
\end{eqnarray}
For $B \to V \ell^-\overline{\nu}_{\ell}$ decay, besides the
branching ratio and the normalized FB asymmetry of charged lepton,
another interesting observable is  the ratio of longitudinal to
transverse polarization of the vector meson $\Gamma_L^V/\Gamma_T^V$,
which can be derived from the following differential expressions
\begin{eqnarray}
\frac{d\Gamma_L^{\spur{R_p}}}{ds}&=&\frac{\sqrt{\lambda_V}}{2^7\pi^3m^3_Bc^2_V}\left(1-\frac{m_\ell^2}{s}\right)^2
\left\{\left|\frac{G_F}{\sqrt{2}}V_{ub}-\sum_i
\frac{\lambda'_{n3i}\lambda'^{*}_{m1i}}
 {8m^2_{\tilde{d}_{iR}}}\right|^2\left(\frac{4}{3}+\frac{2m_\ell^2}{3s}\right) \right.\nonumber\\
 &&\times
  \left(\left[A^V_1(s)\right]^2\frac{(m_B^2-m_V^2-s)^2(m_B+m_V)^2}{4m_V^2}\right.\nonumber\\
 && \left.+\left[A^V_2(s)\right]^2\frac{\lambda_V^2}{4m_V^2(m_B+m_V)^2}
  -A^V_1(s)A^V_2(s)\frac{(m_B^2-m_V^2-s)\lambda_V}{4m_V^2}\right)\nonumber\\
&& \left.+2\left|\frac{G_F}{\sqrt{2}}V_{ub}-\sum_i
\frac{\lambda'_{n3i}\lambda'^{*}_{m1i}}
 {8m^2_{\tilde{d}_{iR}}}+\sum_i\frac{\lambda_{inm}\lambda'^{*}_{i13}}{4m^2_{\tilde{\ell}_{iL}}}
 \frac{s}{m_\ell(\overline{m}_b+\overline{m}_u)}\right|^2
\left[A^V_0(s)\right]^2\frac{m_\ell^2}{s}\lambda_V\right\},\\
\frac{d\Gamma_T^{\spur{R_p}}}{ds}&=&\frac{\sqrt{\lambda_V}}{2^7\pi^3m^3_Bc^2_V}\left(1-\frac{m_\ell^2}{s}\right)^2
\left|\frac{G_F}{\sqrt{2}}V_{ub}-\sum_i
\frac{\lambda'_{n3i}\lambda'^{*}_{m1i}}{8m^2_{\tilde{d}_{iR}}}\right|^2\frac{8}{3}\nonumber\\
&&\times
\left\{\left[A^V_1(s)\right]^2(m^2_\ell+2s)(m_B+m_V)^2+\left[V^V(s)\right]^2\frac{\lambda_V(m^2_\ell+2s)}{(m_B+m_V)^2}
\right\}.
\end{eqnarray}

In this section, we give the expressions of only the exclusive $b\to
u \ell^- \overline{\nu}_\ell$ decays, but we will use the CP
averaged results of the exclusive $b\to u \ell^-
\overline{\nu}_\ell$ and $\overline{b}\to \overline{u} \ell^+
\nu_\ell$ decays in our numerical analysis.

\section{Input Parameters}

The input parameters except the form factors are collected in Table
I. In our numerical results, we will use the input parameters, which
are varied randomly within $1\sigma$ range.
\begin{table}[htbp]
\centerline{\parbox{16cm}{\small Table I: Default values of the
input parameters and the $\pm1 \sigma$ error ranges for the sensitive
parameters used in our numerical calculations.}} \vspace{0.3cm}
\begin{center}
\begin{tabular}{lr}\hline\hline
$m_{_{B_s}}=5.366~{\rm GeV},~~m_{_{B_d}}=5.279~{\rm GeV},~~m_{_{B_u}}=5.279~{\rm GeV},~~m_{_{K^{*\pm}}}=0.892~{\rm GeV},$ \\
$m_{_{\pi^\pm}}=0.140~{\rm GeV},~~m_{_{\pi^0}}=0.135~{\rm
GeV},~~m_{\rho}=0.775~{\rm GeV},$
$m_{_{K^\pm}}=0.494~{\rm GeV},$ \\
$\overline{m}_b(\overline{m}_b)=(4.20\pm0.07)~{\rm
GeV},~~\overline{m}_u(2~{\rm GeV})=0.0015\sim
0.003~{\rm GeV},$ \\
$m_e=0.511\times10^{-3}~{\rm GeV},~~m_\mu=0.106~{\rm GeV},$~~
$m_{\tau}=1.777~{\rm GeV}. $&\cite{PDG2006}\\\hline
$\tau_{_{B_s}}=(1.437^{+0.030}_{-0.031})~ps,~~\tau_{_{B_{d}}}=(1.530\pm
0.009)~ps,~~\tau_{_{B_{u}}}=(1.638\pm
0.011)~ps.$&\cite{PDG2006}\\\hline $f_{_{B_u}}=0.161\pm0.013~{\rm
GeV}.$&\cite{BallZwicky}\\\hline
$|V_{ub}|=(4.31\pm0.39)\times10^{-3}.$&\cite{HFAG}\\\hline
$\epsilon_0\in[-0.01, 0.01].$& \cite{BurasHcorrection}\\\hline
\end{tabular}
\end{center}
\end{table}

For the form factors involving the $B\to P(V)$ transitions, we will
use the recent  light-cone  QCD sum rules (LCSRs)  results
\cite{BallZwicky},
 which are renewed with  radiative corrections to
the leading twist wave functions and SU(3) breaking effects.
 For the $s$-dependence of the form factors,
they can be parameterized in terms of simple formulae with two or
three parameters. The form factors $V^V, A_0^V$ and $f_+^\pi$ are
parameterized by
\begin{eqnarray}
F(s)=\frac{r_1}{1-s/m^2_{R}}+\frac{r_2}{1-s/m^2_{fit}}.\label{r12mRfit}
\end{eqnarray}
For the form factors $A^V_2$ and $f_+^K$, it is more appropriate to
expand to second order around the pole, yielding
\begin{eqnarray}
F(s)=\frac{r_1}{1-s/m^2}+\frac{r_2}{(1-s/m^2)^2},\label{r12mfit}
\end{eqnarray}
where $m=m_{fit}$ for $A_2^V$ and $m=m_{R}$ for $f_+^K$.  The fit
formula
 for $A_1^V$ and $f_0^P$ is
\begin{eqnarray}
F(s)=\frac{r_2}{1-s/m^2_{fit}}.\label{r2mfit}
\end{eqnarray}
However, $B_s\to K$ form factors are not given in LCSR results
\cite{BallZwicky}. After discussions with authors of Ref. \cite{BallZwicky}, we  obtain them as
\begin{eqnarray}
F^{B_s\to K}(s)=F^{B_{u,d}\to K}(s)\left(\frac{F^{B_s\to
K^*}(s)}{F^{B_{u,d}\to K^*}(s)}\right).
\end{eqnarray}
All the
corresponding  parameters for these form factors are collected in
Table II.
\begin{table}[htb]
\centerline{\parbox{13cm}{\small Table II: Fit for form factors
involving the $B\to K^{(*)}$ and $B\to \rho(\pi)$ transitions valid
for general $s$ \cite{BallZwicky}.}} \vspace{0.3cm}
\begin{center}
\begin{tabular}{ccccccc}\hline\hline
$F(s)$&$~F(0)~$&$~r_1~$&$~m_R^2~$&$~r_2~$&$~m^2_{fit}~$&~fit
Eq.\\\hline $V^{B_{u,d}\to
\rho}$&$0.323\pm 0.030$&$1.045$&$5.32^2$&$-0.721$&$38.34$&(\ref{r12mRfit})\\
$A_0^{B_{u,d}\to
\rho}$&$0.303\pm 0.029$&$1.527$&$5.28^2$&$-1.220$&$33.36$&(\ref{r12mRfit})\\
$A_1^{B_{u,d}\to
\rho}$&$0.242\pm 0.023$&$$&$$&$0.240$&$37.51$&(\ref{r2mfit})\\
$A_2^{B_{u,d}\to
\rho}$&$0.221\pm 0.023$&$0.009$&$$&$0.212$&$40.82$&(\ref{r12mfit})\\\hline
$V^{B_{u,d}\to
K^*}$&$0.411\pm 0.033$&$0.923$&$5.32^2$&$-0.511$&$49.40$&(\ref{r12mRfit})\\
$A_0^{B_{u,d}\to
K^*}$&$0.374\pm 0.033$&$1.364$&$5.28^2$&$-0.990$&$36.78$&(\ref{r12mRfit})\\
$A_1^{B_{u,d}\to
K^*}$&$0.292\pm 0.028$&$$&$$&$0.290$&$40.38$&(\ref{r2mfit})\\
$A_2^{B_{u,d}\to
K^*}$&$0.259\pm 0.027$&$-0.084$&$$&$0.342$&$52.00$&(\ref{r12mfit})\\\hline
$V^{B_s\to
K^*}$&$0.311\pm 0.026$&$2.351$&$5.42^2$&$-2.039$&$33.10$&(\ref{r12mRfit})\\
$A_0^{B_s\to
K^*}$&$0.360\pm 0.034$&$2.813$&$5.37^2$&$-2.509$&$31.58$&(\ref{r12mRfit})\\
$A_1^{B_s\to
K^*}$&$0.233\pm 0.022$&$$&$$&$0.231$&$32.94$&(\ref{r2mfit})\\
$A_2^{B_s\to
K^*}$&$0.181\pm 0.025$&$-0.011$&$$&$0.192$&$40.14$&(\ref{r12mfit})\\\hline
$f^{B_{u,d}\to
\pi}_+$&$0.258\pm 0.031$&$0.744$&$5.32^2$&$-0.486$&$40.73$&(\ref{r12mRfit})\\
$f^{B_{u,d}\to
\pi}_0$&$0.258\pm 0.031$&$0$&$$&$0.258$&$33.81$&(\ref{r2mfit})\\\hline
$f^{B_{u,d}\to
K}_+$&$0.331\pm 0.041$&$0.162$&$5.41^2$&$0.173$&$$&(\ref{r12mfit})\\
$f^{B_{u,d}\to
K}_0$&$0.331\pm 0.041$&$0$&$$&$0.331$&$37.46$&(\ref{r2mfit})\\\hline
\end{tabular}
\end{center}
\end{table}

We have several remarks on the input parameters:
\begin{itemize}
\item \underline{Form factor}:
The uncertainties of form factors at $s=0$ induced by $F(0)$  are
considered.

\item \underline{CKM matrix element}: Using experimental measurements of $|V_{ub}|$ from the inclusive $b\to
u$ semileptonic $B$ decays, these exclusive $b\to u
\ell^-\overline{\nu}_\ell$ decays can be used to constrain the
parameters of theories beyond the SM. The weak phase $\gamma$ is
well constrained in the SM,  however, with the presence of $R_p$
violation, this constraint may be relaxed. We will not take $\gamma$
within the SM range, but vary it randomly in the range of 0 to $\pi$
to obtain conservative limits on RPV coupling products.

\item  \underline{RPV coupling}:  When we study the RPV effects, we  consider
only one RPV coupling product contributes at one time, neglecting
the interferences between different RPV coupling products, but
keeping their interferences  with the SM amplitude. We assume the
masses of sfermion are 100 GeV. For other values of the sfermion
masses, the bounds on the couplings in this paper can be easily
obtained by scaling them by factor
$\tilde{f}^2\equiv(\frac{m_{\tilde{f}}}{100~\rm{GeV}})^2$.
\end{itemize}

\section{Numerical results in the MSSM with RPC}

In this section, we study the charged Higgs contributions to
the exclusive $\bar{b}\to \bar{u} \ell^+ \nu_{\ell}$ decays in the
MSSM with RPC. Since the couplings of the charged Higgs to the
leptons are always proportional to the charged lepton masses (see the
foregoing equations), it is easily to understand that the effects of
the charged Higgs will not significantly affect in the case of the light leptonic
decays, so we  only present the charged Higgs contributions to the
exclusive $\bar{b}\to \bar{u} \tau^+ \nu_{\tau}$ decays. Based on
the constraint of the charged Higgs effects from the measurement on
$\mathcal{B}(B^+\to\tau^+\nu_\tau)$, we investigate these effects on
$\mathcal{B},~d\mathcal{B}/ds$, $\overline{\mathcal{A}}_{FB}$ and
$\Gamma^V_L/\Gamma^V_T$ in the exclusive  $\bar{b}\to \bar{u} \tau^+
\nu_{\tau}$ semileptonic decays.

 Note that the charged Higgs effects on
the exclusive $\bar{b}\to \bar{u} \tau^+ \nu_{\tau}$ decays have
been discussed  in Ref. \cite{Higgs Gengchaoqiang},  which fixed
$\tan\beta=50$ and let physical quantity as a function of $m_H$.
Here we will not choose $\tan\beta$ as a fixed value  but let
observable as a function of $\tan\beta$ and $m_H$ to study the
effects of $\tan\beta$ and $m_H$. In addition, we will investigate
the charged Higgs contributions to $\Gamma^V_L/\Gamma^V_T$, which
has not been studied yet. For the exclusive $\bar{b} \to \bar{u}
\tau^+ \nu_{\tau}$ decays, the purely leptonic decay $B^+_u\to
\tau^+\nu_\tau$ has been measured by \textit{B{\footnotesize
A}B{\footnotesize AR}} \cite{BABARBtotaunu} and Belle
\cite{BelleBtotaunu}. We will use the averaged experimental data
from Heavy Flavor Averaging Group  \cite{HFAG}
\begin{eqnarray}
\mathcal{B}(B^+_u\to
\tau^+\nu_\tau)=(1.41^{+0.43}_{-0.42})\times10^{-4}.\label{tauBranching}
\end{eqnarray}

\begin{figure}[t]
\begin{center}
\includegraphics[scale=0.8]{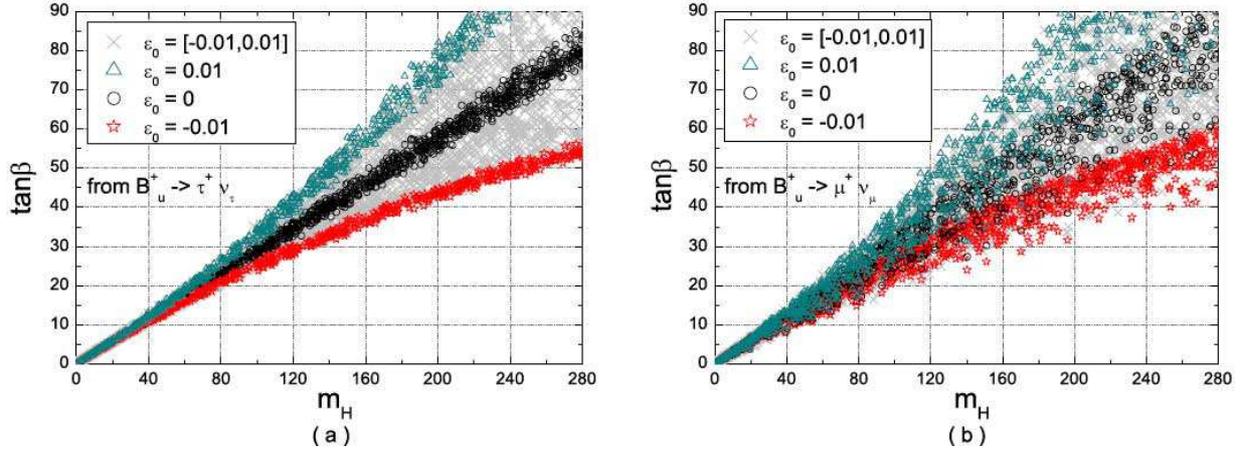}
\end{center}\vspace{-1cm}
\caption{The allowed regions in the $\tan\beta$-$m_{H}$ plane for
different values of $\epsilon_0$. Plot (a) is constrained from the
experimental date of $\mathcal{B}(B^+_u\to\tau^+\nu_\tau)$, and plot
(b) is constrained from the upper limit of
$\mathcal{B}(B^+_u\to\mu^+\nu_\mu)$.}\label{Hbounds}
\end{figure}

Using the experimental data of $\mathcal{B}(B^+_u\to
\tau^+\nu_\tau)$ varied randomly within $1\sigma$ range and
considering the theoretical uncertainties, we constrain the allowed
range of $\tan\beta/m_H$, which is shown in Fig. \ref{Hbounds}(a).
The corresponding bound from  the upper limit of
$\mathcal{B}(B^+_u\to\mu^+\nu_\mu)<1.7\times10^{-6}$ is also
displayed in Fig. \ref{Hbounds}(b), in which the bound is weaker
than one from $\mathcal{B}(B^+_u\to \tau^+\nu_\tau)$. At present,
the most stringent bound comes from $B^+_u\to\tau^+\nu_\tau$. The
numerical ranges of $\tan\beta/m_H$ without the radiative
corrections ($\epsilon_0=0$) and with inclusion of radiative
corrections ($\epsilon_0\in[-0.01,0.01]$) are given in Table III. In
Ref. \cite{HiggsAkeroyd}, from the experimental upper limit of
$\mathcal{B}(B^+_u\to \tau^+\nu_\tau)<4.1\times10^{-4}$, the authors
got $\tan\beta/m_H=0.34(0.36,0.32)~{\rm GeV}^{-1}$ for
$f_{_{B_u}}=0.2(0.17,0.23)~{\rm GeV}$ with $\epsilon_0=0$. Our
bounds on $\tan\beta/m_H$, from new data of $\mathcal{B}(B^+_u\to
\tau^+\nu_\tau)$ and with considering all theoretical uncertainties,
are much stronger than theirs, as shown in Table III.

\begin{table}[t]
\centerline{\parbox{14cm}{\small Table III: The allowed ranges of
$\tan\beta/m_H$ from $\mathcal{B}(B^+_u\to\tau^+\nu_{\tau})$ and
$\mathcal{B}(B^+_u\to\mu^+\nu_{\mu})$.}} \vspace{0.3cm}
\begin{center}
\begin{tabular}{c|c|c}\hline\hline
&$\epsilon_0=0$&~~$\epsilon_0\in[-0.01,0.01]$\\\hline
 $\tan\beta/m_H$ from
$\mathcal{B}(B^+_u\to\tau^+\nu_{\tau})$&~~[0.26,0.31]$~{\rm
GeV}^{-1}$~~&~~[0.18,0.49]$~{\rm GeV}^{-1}$\\\hline $\tan\beta/m_H$
from $\mathcal{B}(B^+_u\to\mu^+\nu_{\mu})$&~~[0.20,0.34]$~{\rm
GeV}^{-1}$~~&~~[0.15,0.57]$~{\rm GeV}^{-1}$\\\hline
\end{tabular}
\end{center}
\end{table}

Using the constrained $\tan\beta/m_H$ from $\mathcal{B}(B^+_u\to
\tau^+\nu_\tau)$, one can predict the charged Higgs effects on the
semileptonic decays
 $B^+_u\to\pi^0\tau^+\nu_\tau$,
 $B^0_d\to\pi^-\tau^+\nu_\tau$, $B^0_s\to K^- \tau^+\nu_\tau$,  $B^+_u\to\rho^0\tau^+\nu_\tau$,
 $B^0_d\to\rho^-\tau^+\nu_\tau$ and $B^0_s\to K^{*-} \tau^+\nu_\tau$. With the
expressions for  $\mathcal{B}$ and $\Gamma_L^V/\Gamma_T^V$ at hand,
we perform a scan on the input parameters and the newly constrained
$\tan\beta/m_H$. Then, the allowed ranges for $\mathcal{B}$ and
$\Gamma_L^V/\Gamma_T^V$ are obtained including the charged Higgs
contributions, which satisfy present experimental constraint of
$\mathcal{B}(B^+_u\to \tau^+\nu_\tau)$ shown in Eq.
(\ref{tauBranching}). Our numerical results are summarized  in Table VI, in which we find that the charged
Higgs contributions could slightly reduce $\mathcal{B}(B\to P(V)
\tau \nu_\tau)$ and $\frac{\Gamma_L}{\Gamma_T}(B\to V
\tau^+\nu_\tau)$.

\begin{table}[htbp]
\centerline{\parbox{12cm}{\small Table VI: The theoretical
predictions of the exclusive $\overline{b} \to \overline{u} \tau^+
\nu_{\tau}$ decays for $\mathcal{B}(\times10^{-4})$ and
$\Gamma_L^V/\Gamma_T^V$ in the SM and in the MSSM with RPC.}}
\vspace{0.3cm}
\begin{center}\small{
\begin{tabular}{l|c|c}\hline\hline
&SM value &MSSM value w/ RPC\\\hline $\mathcal{B}(B^+_u\to\pi^0
\tau^+\nu_\tau)$&$[0.58,1.22]$&$[0.43,0.96]$\\\hline
$\mathcal{B}(B^0_d\to\pi^-
\tau^+\nu_\tau)$&$[1.12,2.28]$&$[0.80,1.79]$\\\hline
$\mathcal{B}(B^0_s\to K^-
\tau^+\nu_\tau)$&$[1.47,3.05]$&$[1.02,2.37]$\\\hline
$\mathcal{B}(B^+_u\to\rho^0
\tau^+\nu_\tau)$&$[0.97,2.19]$&$[0.83,2.02]$\\\hline
$\mathcal{B}(B^0_d\to\rho^-
\tau^+\nu_\tau)$&$[1.83,4.08]$&$[1.56,3.78]$\\\hline
$\mathcal{B}(B^0_s\to K^{*-}
\tau^+\nu_\tau)$&$[2.08,4.46]$&$[1.64,4.06]$\\\hline
$\frac{\Gamma_L}{\Gamma_T}(B^+_u\to\rho^0
\tau^+\nu_\tau)$&$[0.65,1.19]$&$[0.45,1.03]$\\\hline
$\frac{\Gamma_L}{\Gamma_T}(B^0_d\to\rho^-
\tau^+\nu_\tau)$&$[0.65,1.19]$&$[0.45,1.03]$\\\hline
$\frac{\Gamma_L}{\Gamma_T}(B^0_s\to K^{*-}
\tau^+\nu_\tau)$&$[0.84,1.38]$&$[0.58,1.11]$\\\hline
\end{tabular}}
\end{center}
\end{table}

\begin{figure}[t]
\begin{center}
\includegraphics[scale=0.82]{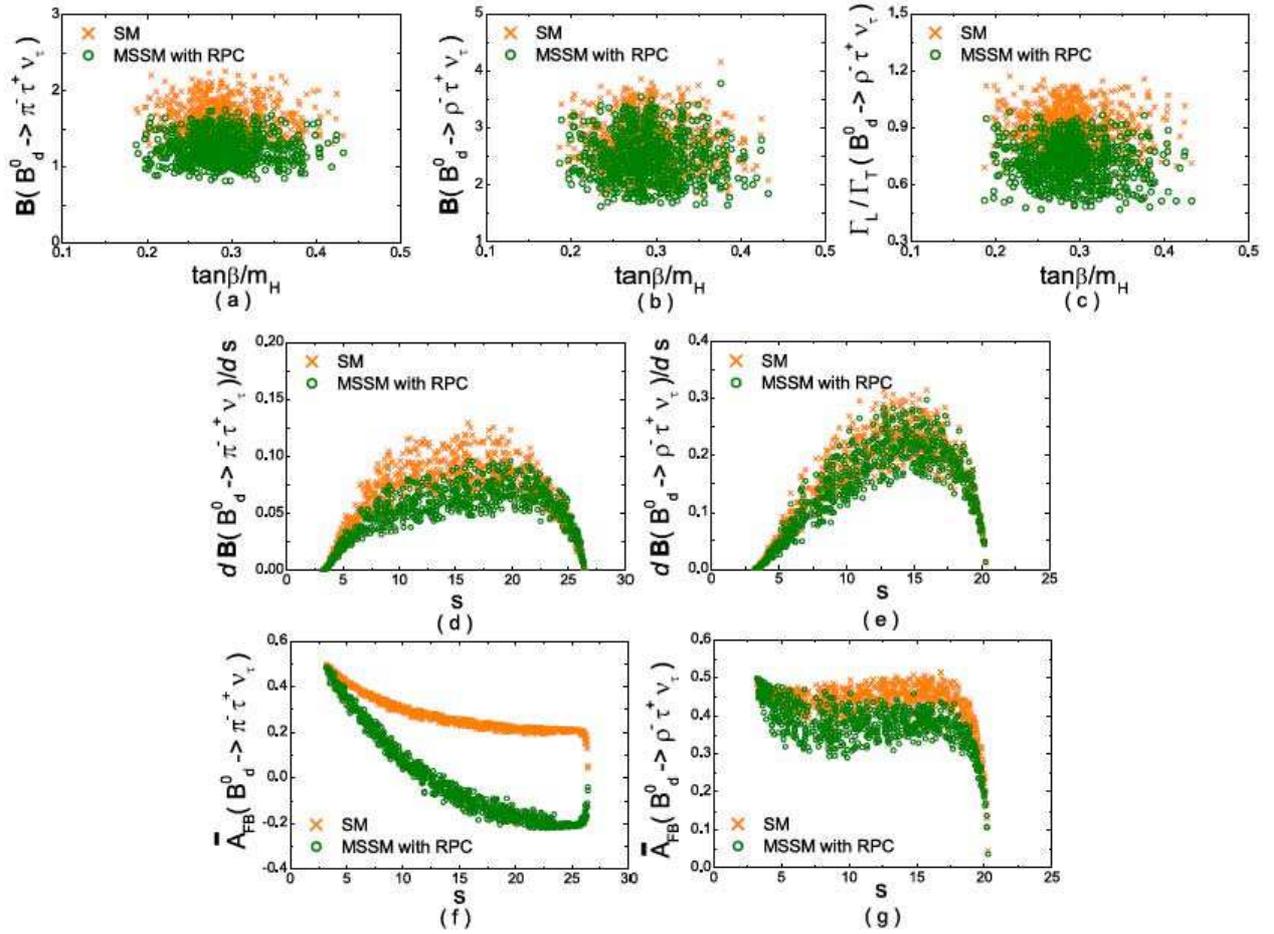}
\end{center}\vspace{-1cm}
\caption{The charged Higgs effects on $B^0_d\to \pi^-(\rho^-) \tau^+
 \nu_\tau$ decays in the MSSM with RPC.
 $\mathcal{B}$ and $d\mathcal{B}/ds$  are in unit of $10^{-4}$.}\label{Htau2D}
\end{figure}

Now, we present correlations between the physical observables  and
the charged Higgs effects by the two-dimensional scatter plots, and
moreover, we give the SM predictions for comparison. The charged
Higgs effects on $B^+_u\to \pi^0 \tau^+ \nu_\tau$, $B^0_d\to \pi^-
\tau^+ \nu_\tau$ and $B^0_s\to K^- \tau^+ \nu_\tau$ are very similar
to each other, therefore we will  take $B^0_d\to \pi^- \tau^+
\nu_\tau$ decay as an example. For the same reason, we will only
display the charged Higgs effects on $B^0_d\to \rho^- \tau^+
\nu_\tau$ among other three decay modes $B^+_u\to \rho^0 \tau^+
\nu_\tau$, $B^0_d\to\rho^- \tau^+ \nu_\tau$ and $B^0_s\to K^{*-}
\tau^+ \nu_\tau$. The charged Higgs effects on $B^0_d\to
\pi^-(\rho^-) \tau^+
 \nu_\tau$ decays are shown in Fig. \ref{Htau2D}.

{}From Fig. \ref{Htau2D}(a-c), we can see that
$\mathcal{B}(B^0_d\to\pi^-\tau^+\nu_\tau)$,
 $\mathcal{B}(B^0_d\to\rho^-\tau^+\nu_\tau)$ and
 $\frac{\Gamma_L}{\Gamma_T}(B^0_d\to\rho^-\tau^+\nu_\tau)$ are not much
 sensitive
 to the change of  $\tan\beta/m_H$, but the charged Higgs contributions can slightly reduce these quantities.
 As shown in Fig. \ref{Htau2D}(d-g), the charged Higgs
 have also reducing  effects on $d\mathcal{B}/ds$ and
$\overline{\mathcal{A}}_{FB}$. Especially, the sign of
 $\overline{\mathcal{A}}_{FB}(B^0_d\to\pi^-\tau^+\nu_\tau)$ could be
changed by the effect. According to Eqs.
(\ref{brBtoPlnu})-(\ref{NAFBP}),  since the normalized FB asymmetry
of $B\to P\ell^+\nu_\ell$ is associated with $m^2_\ell
f^P_{0}(s)f^P_{+}(s)$ and not suppressed by $s$, we can easily
understand that $\overline{\mathcal{A}}_{FB}(B^0_d\to
\pi^-\tau^+\nu_\tau)$ shown in Fig. \ref{Htau2D}(f) could be
significantly affected by the charged Higgs couplings. Therefore,
$\overline{\mathcal{A}}_{FB}(B\to P\tau^+\nu_\tau)$ are very
powerful quantities to be measured, to constrain the charged Higgs
effects in the MSSM with RPC.

\section{Numerical results  in the RPV MSSM}

\subsection{The exclusive $\bar{b} \to \bar{u} \tau^+ \nu_{\tau}$ decays}

\begin{figure}[t]
\begin{center}
\includegraphics[scale=0.6]{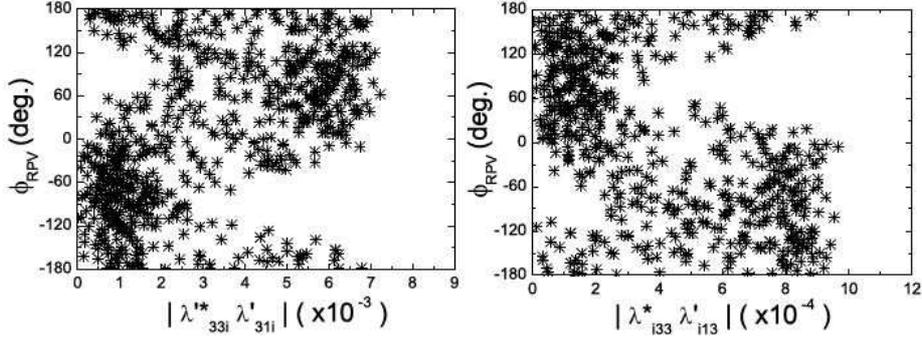}
\end{center}\vspace{-1cm}
\caption{The allowed parameter spaces for the relevant RPV coupling
products constrained by the experimental data of
$\mathcal{B}(B^+_u\to \tau^+\nu_\tau)$.}\label{taubounds}
\end{figure}

There are two RPV coupling products,
$\lambda'^*_{33i}\lambda'_{31i}$ and
$\lambda^*_{i33}\lambda'_{i13}$,  contributing to seven exclusive
$\overline{b} \to \overline{u} \tau^+ \nu_{\tau}$ decay modes,
 $B^+_u\to\tau^+\nu_\tau$, $B^+_u\to\pi^0\tau^+\nu_\tau$,
 $B^0_d\to\pi^-\tau^+\nu_\tau$, $B^0_s\to K^- \tau^+\nu_\tau$,  $B^+_u\to\rho^0\tau^+\nu_\tau$,
 $B^0_d\to\rho^-\tau^+\nu_\tau$ and $B^0_s\to K^{*-} \tau^+\nu_\tau$.
 We use the experimental data of $\mathcal{B}(B^+_u\to
\tau^+\nu_\tau)$, which is varied randomly within $1\sigma$ range to
constrain the two RPV coupling products. Our bounds on the two RPV
coupling products are demonstrated in Fig. \ref{taubounds}, in which we find
that every RPV weak phase is not much constrained, but the
modulus of the relevant RPV coupling products can be tightly upper
limited.
The upper limits for the relevant RPV coupling products are
summarized in Table V.
%For comparison,  the existing bounds are also listed.
Note that the  bounds on the direct quadric
couplings have not been estimated in previous $\bar{b} \to \bar{u} \tau^+ \nu_{\tau}$ studies.
Our bounds on the RPV quadric couplings from $B^+_u\to\tau^+
\nu_\tau$ are weaker than the bounds, which are calculated from the products of the smallest values of
two single  couplings in \cite{0406029,0406039}.

\begin{table}[t]
\centerline{\parbox{9cm}{Table V: \small {Bounds for  the relevant
RPV coupling products by $B^+_u \to \tau^+ \nu_\tau$ decay for 100
${\rm GeV}$ sfermions. }}} \vspace{0.5cm}
\begin{center}
\begin{tabular}{c|c}\hline\hline
Couplings&~Bounds~~~~~~~[Processes] \\
\hline $|\lambda'^*_{33i}\lambda'_{31i}|$&$\leq7.28\times
10^{-3}~[B^+_u \to \tau^+ \nu_\tau]$ \\
$|\lambda^*_{i33}\lambda'_{i13}|$&$\leq9.65\times 10^{-4}~[B^+_u \to
\tau^+ \nu_\tau]$ \\\hline
\end{tabular}
\end{center}
\end{table}

Using the constrained  parameter spaces shown in Fig.
\ref{taubounds}, we will  predict the RPV effects on  other
quantities which have not been measured yet in the exclusive
$\overline{b} \to \overline{u} \tau^+ \nu_{\tau}$ decays. The
allowed ranges for $\mathcal{B}$ and $\Gamma_L^V/\Gamma_T^V$ are
obtained with the different RPV coupling products, which are
summarized  in Table VI.
\begin{table}[t]
\centerline{\parbox{15.2cm}{\small Table VI: The theoretical
predictions of the exclusive $\overline{b} \to \overline{u} \tau^+
\nu_{\tau}$ decays for $\mathcal{B}(\times10^{-4})$ and
$\Gamma_L^V/\Gamma_T^V$ in the SM and the RPV MSSM. The RPV MSSM
predictions are obtained by the constrained regions of the different
RPV coupling products. }} \vspace{0.3cm}
\begin{center}\small{
\begin{tabular}{l|c|c|c}\hline\hline
&SM value & MSSM value w/ $\lambda'^*_{33i}
 \lambda'_{31i}$& MSSM value w/  $\lambda^*_{i33}
 \lambda'_{i13}$\\\hline
$\mathcal{B}(B^+_u\to\pi^0
\tau^+\nu_\tau)$&$[0.58,1.22]$&$[0.78,2.47]$&$[0.49,1.30]$\\\hline
$\mathcal{B}(B^0_d\to\pi^-
\tau^+\nu_\tau)$&$[1.12,2.28]$&$[1.45,4.59]$&$[0.91,2.41]$\\\hline
$\mathcal{B}(B^0_s\to K^-
\tau^+\nu_\tau)$&$[1.47,3.05]$&$[1.92,5.91]$&$[1.18,3.35]$\\\hline
$\mathcal{B}(B^+_u\to\rho^0
\tau^+\nu_\tau)$&$[0.97,2.19]$&$[1.42,4.07]$&$[0.89,2.17]$\\\hline
$\mathcal{B}(B^0_d\to\rho^-
\tau^+\nu_\tau)$&$[1.83,4.08]$&$[2.64,7.57]$&$[1.65,4.04]$\\\hline
$\mathcal{B}(B^0_s\to K^{*-}
\tau^+\nu_\tau)$&$[2.08,4.46]$&$[2.85,9.62]$&$[1.96,4.57]$\\\hline
$\frac{\Gamma_L}{\Gamma_T}(B^+_u\to\rho^0
\tau^+\nu_\tau)$&$[0.65,1.19]$&$\cdots\cdots$&$[0.47,1.22]$\\\hline
$\frac{\Gamma_L}{\Gamma_T}(B^0_d\to\rho^-
\tau^+\nu_\tau)$&$[0.65,1.19]$&$\cdots\cdots$&$[0.47,1.22]$\\\hline
$\frac{\Gamma_L}{\Gamma_T}(B^0_s\to K^{*-}
\tau^+\nu_\tau)$&$[0.84,1.38]$&$\cdots\cdots$&$[0.68,1.41]$\\\hline
\end{tabular}}
\end{center}
\end{table}
We can find some salient features of the numerical results listed in
Table VI.

\begin{itemize}
\item[\textcircled{\scriptsize 1}]
The contributions of $\lambda'^*_{33i}\lambda'_{31i}$ due to squark
exchange will little enhance the branching ratios $\mathcal{B}(B\to
P \tau^+ \nu_\tau)$ and $\mathcal{B}(B\to V \tau^+ \nu_\tau)$.
Because the effective Hamiltonian of squark exchange is proportional
to operator
$(\bar{b}\gamma_\mu(1-\gamma_5)u)(\overline{\nu}_{\tau}\gamma^\mu(1-\gamma_5)\tau)$,
which is the same as the SM one, the effects of squark exchange are
completely canceled in $\frac{\Gamma_L}{\Gamma_T}(B\to V
\tau^+\nu_\tau)$.

\item[\textcircled{\scriptsize 2}] As for the contributions  of  $\lambda^*_{i33}
 \lambda'_{i13}$ due to slepton exchange, the slepton
exchange coupling has not obvious effects on $\mathcal{B}(B\to P(V)
\tau^+ \nu_\tau)$, but the allowed ranges of
$\frac{\Gamma_L}{\Gamma_T}(B\to V \tau^+\nu_\tau)$ can be enlarged
by this coupling, especially, their allowed lower limits are
observably decreased.
\end{itemize}

\begin{figure}[t]
\begin{center}
\includegraphics[scale=0.41]{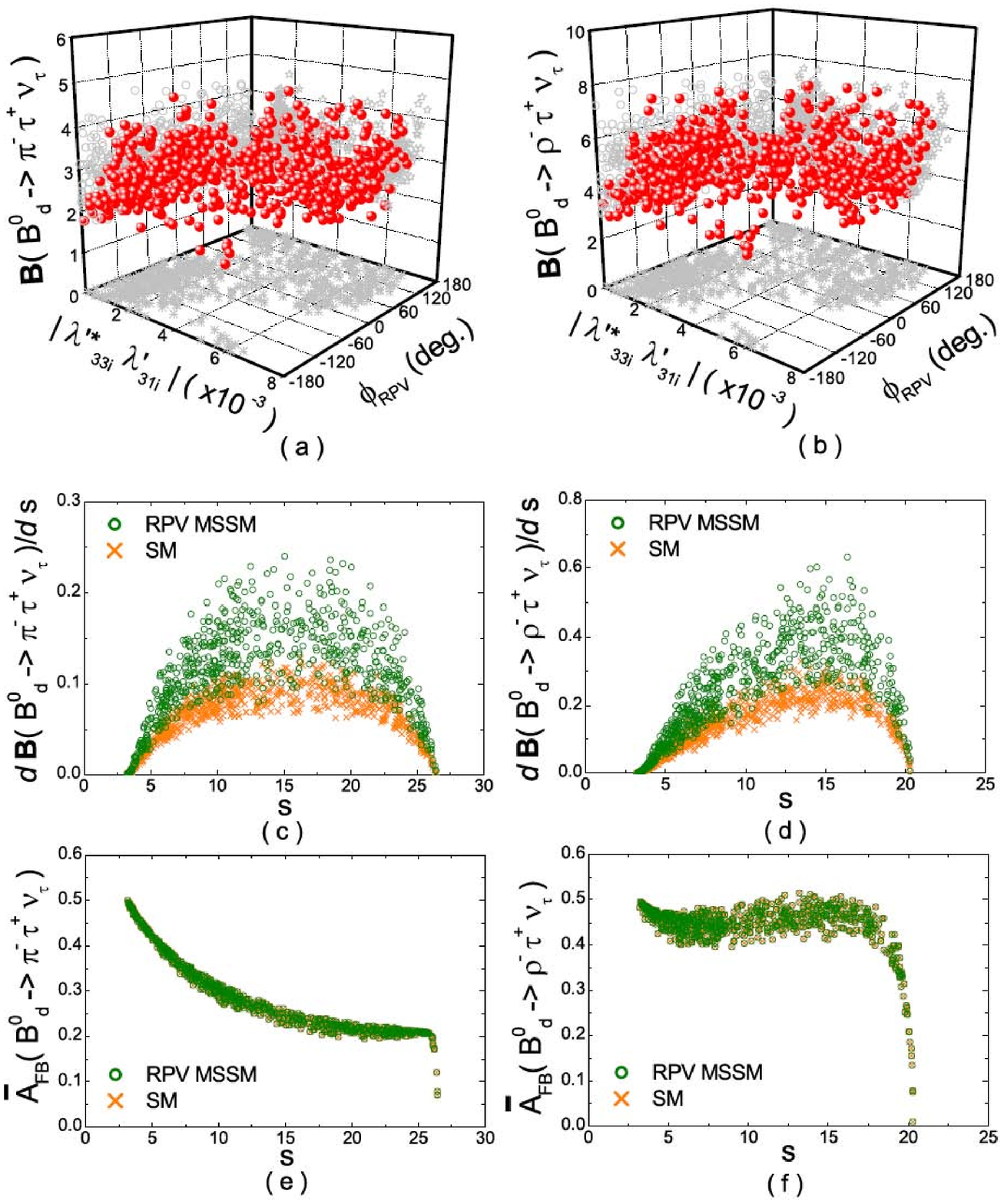}
\end{center}\vspace{-1cm}
\caption{ The effects of RPV coupling $\lambda'^*_{33i}
 \lambda'_{31i}$ on $B^0_d\to \pi^-(\rho^-) \tau^+
 \nu_\tau$ decays.
 $\mathcal{B}$ and $d\mathcal{B}/ds$  are in unit of $10^{-4}$.}\label{figlpslptau}
%\end{figure}
%\begin{figure}[b]
\begin{center}
\includegraphics[scale=0.6]{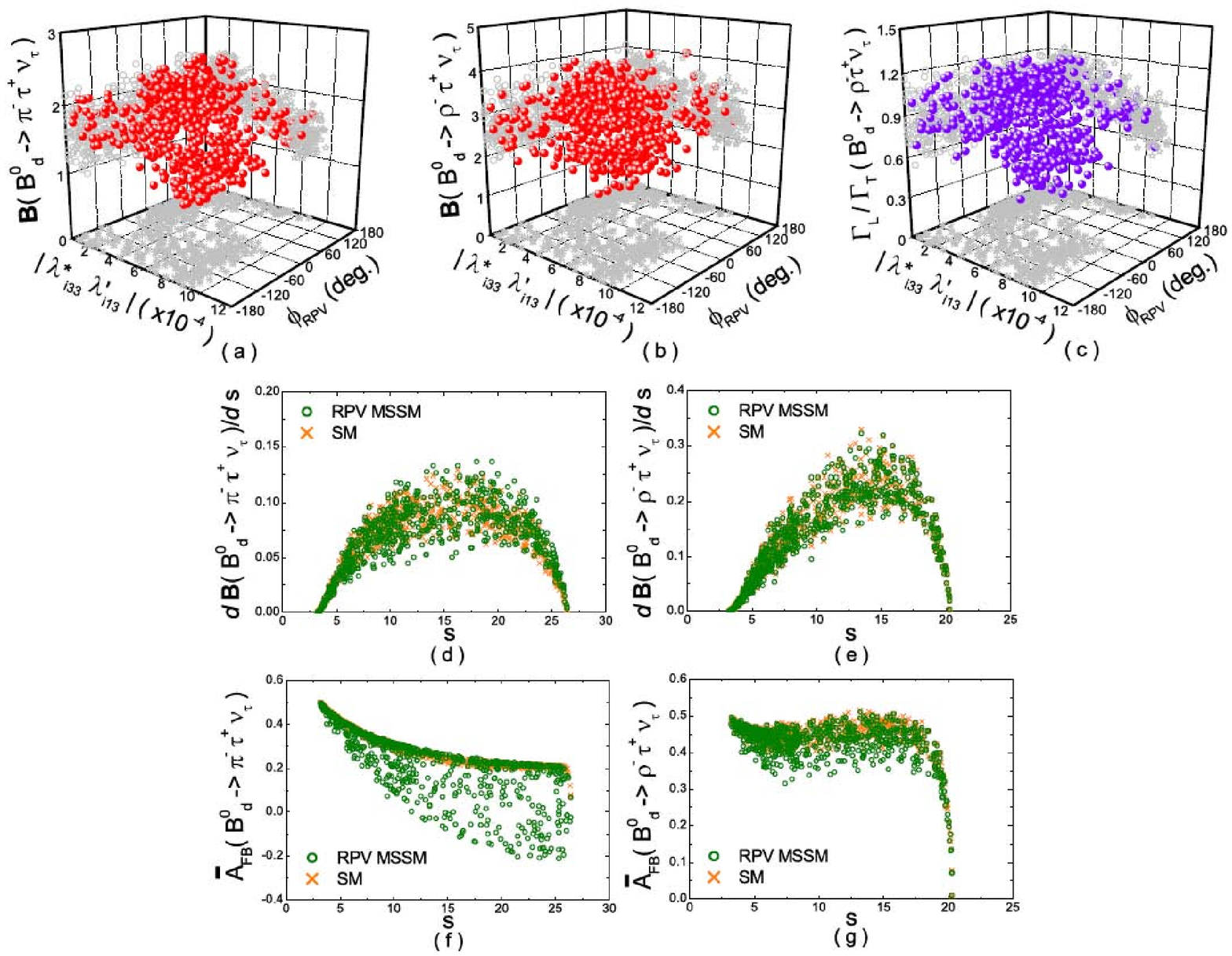}
\end{center}\vspace{-1cm}
\caption{The effects of RPV coupling $\lambda^*_{i33}
 \lambda'_{i13}$ on $B^0_d\to \pi^-(\rho^-) \tau^+
 \nu_\tau$ decays.
 $\mathcal{B}$ and $d\mathcal{B}/ds$  are in unit of $10^{-4}$.}\label{figlslptau}
\end{figure}

For each RPV coupling product, we can present the correlations of
$\mathcal{B}$ and $\Gamma_L^V/\Gamma_T^V$ within the constrained
parameter space displayed in Fig. \ref{taubounds} by the
three-dimensional scatter plots. The differential branching ratio
$d\mathcal{B}/ds$ and the normalized FB asymmetry
$\overline{\mathcal{A}}_{FB}$ can be shown by the two-dimensional
scatter plots. The RPV coupling $\lambda'^*_{33i}\lambda'_{31i}$ or
$\lambda^*_{i33}
 \lambda'_{i13}$ contributions to $B^+_u\to
\pi^0(\rho^0) \tau^+ \nu_\tau$, $B^0_d\to \pi^-(\rho^-) \tau^+
\nu_\tau$ and $B^0_s\to K^-(K^{*-}) \tau^+ \nu_\tau$ are also very
similar to each other.
 So  we will  take an example for  $B^0_d\to \pi^-(\rho^-) \tau^+ \nu_\tau$ decay
  to illustrate the RPV coupling effects.
 The effects of the RPV couplings
 $\lambda'^*_{33i}\lambda'_{31i}$ and $\lambda^*_{i33}
 \lambda'_{i13}$ on $B^0_d\to\pi^-
(\rho^-) \tau^+ \nu_\tau$ decays are shown in Fig. \ref{figlpslptau}
and Fig. \ref{figlslptau}, respectively.

Now we turn to discuss plots of Fig. \ref{figlpslptau} in detail.
The three-dimensional scatter plots Figs. \ref{figlpslptau}(a-b)
show $\mathcal{B}(B^0_d\to\pi^- (\rho^-) \tau^+\nu_\tau)$ correlated
with $|\lambda'^*_{33i}\lambda'_{31i}|$ and its phase
$\phi_{\spur{R_p}}$.  We also give projections to three
perpendicular
 planes, where the $|\lambda'^*_{33i}\lambda'_{31i}|$-$\phi_{\spur{R_p}}$ plane displays the
 constrained
 regions of $\lambda'^*_{33i}\lambda'_{31i}$, as the first plot of Fig. \ref{taubounds}.
 It's shown that $\mathcal{B}(B^0_d\to\pi^-(\rho^-)\tau^+\nu_\tau)$ has some sensitivity to $|\lambda'^*_{33i}\lambda'_{31i}|$
 on the $\mathcal{B}(B^0_d\to\pi^-(\rho^-)\tau^+\nu_\tau)$-$|\lambda'^*_{33i}\lambda'_{31i}|$ plane.
 However, from the $\mathcal{B}(B^0_d\to\pi^-(\rho^-)\tau^+\nu_\tau)$-$\phi_{\spur{R_p}}$ plane,
 we see that $\mathcal{B}(B^0_d\to\pi^-(\rho^-)\tau^+\nu_\tau)$ is very insensitive to  $|\phi_{\spur{R_p}}|$.
 As shown in Fig. \ref{figlpslptau}(e-f),
 $\overline{\mathcal{A}}_{FB}(B^0_d\to\pi^-\tau^+\nu_\tau)$ and
 $\overline{\mathcal{A}}_{FB}(B^0_d\to\rho^-\tau^+\nu_\tau)$ are not obviously
 affected by squark exchange coupling
 $\lambda'^*_{33i}\lambda'_{31i}$, too.
 In Fig. \ref{figlpslptau}(c-d), the $\lambda'^*_{33i}\lambda'_{31i}$
 contributions to $d\mathcal{B}(B^0_d\to\pi^-(\rho^-)\tau^+\nu_\tau)/ds$
 are possibly distinguishable from the SM expectations at all $s$ regions.

Fig. \ref{figlslptau} illustrates the $\lambda^*_{i33}
 \lambda'_{i13}$  contributions to  $B^0_d\to
\pi^-(\rho^-) \tau^+
 \nu_\tau$ decays. $\mathcal{B}(B^0_d\to\pi^-\tau^+\nu_\tau)$,
 $\mathcal{B}(B^0_d\to\rho^-\tau^+\nu_\tau)$ and
 $\frac{\Gamma_L}{\Gamma_T}(B^0_d\to\rho^-\tau^+\nu_\tau)$ are all decreasing
 with $|\lambda^*_{i33}\lambda'_{i13}|$, as shown in
 Fig. \ref{figlslptau}(a-c).  {}From Fig. \ref{figlslptau}(f-g), the effect of $\lambda^*_{i33}
 \lambda'_{i13}$ could allow that $\overline{\mathcal{A}}_{FB}(B^0_d\to\pi^-\tau^+\nu_\tau)$ and
   $\overline{\mathcal{A}}_{FB}(B^0_d\to\rho^-\tau^+\nu_\tau)$ have smaller
   values and, especially, the sign of
   $\overline{\mathcal{A}}_{FB}(B^0_d\to\pi^-\tau^+\nu_\tau)$ could be changed by the effect.
There is similar reason for significant effects of slepton exchange
on $\overline{\mathcal{A}}_{FB}(B\to P\tau^+\nu_\tau)$ as Fig. \ref{Htau2D}(f), $i.e.$ the
normalized FB asymmetry is not suppressed by $m_\ell^2$ and $s$. The
different effects  between the charged Higgs and slepton exchange on
$\overline{\mathcal{A}}_{FB}(B\to P\tau^+\nu_\tau)$,  shown
in Fig. \ref{Htau2D}(f) and Fig. \ref{figlslptau}(f),
come from the RPV weak phase $\phi_{\spur{R_p}}$ and the CKM weak phase $\gamma$. The weak
phases contribute only to the RPV MSSM predictions of
$\overline{\mathcal{A}}_{FB}(B\to
 P\tau^+\nu_\tau)$.

\subsection{The exclusive $b\to u \ell' \nu_{\ell'}~~(\ell'=\mu~\mbox{or}~e)$ decays}

\begin{table}[b]
\centerline{\parbox{16.2cm}{\small Table VII: The experimental data
for the exclusive $\overline{b}\to \overline{u} \ell'^+ \nu_{\ell'}$
decays
\cite{PDG2006,BelleBMlnu,Babarpimlnu1,Babarpimlnu2,BabarM0mlnu,CLEOrholnu,CLEOMmlnu}
and corresponding SM predictions.}} \vspace{0.4cm}
\begin{center}
\begin{tabular}
{l|l|l|l}\hline\hline & Experimental data & SM value for
$\ell'=\mu$&SM value for $\ell'=e$\\\hline $\mathcal{B}(B^+_u\to
\mu^+\nu_\mu)$&$<1.7\times10^{-6}$ 90\%
C.L.&$[2.69,5.30]\times10^{-7}$&\\\hline $\mathcal{B}(B^+_u\to
e^+\nu_e)$&$<9.8\times10^{-7}$ 90\%
C.L.&&$[6.28,12.46]\times10^{-12}$\\\hline $\mathcal{B}(B^+_u\to
\pi^0\ell'^+\nu_{\ell'})$&$(0.75\pm0.09)\times10^{-4}$&$[0.76,1.75]\times10^{-4}$&$[0.75,1.75]\times10^{-4}$\\\hline
$\mathcal{B}(B^0_d\to
\pi^-\ell'^+\nu_{\ell'})$&$(1.41\pm0.08)\times10^{-4}$&$[1.41,3.25]\times10^{-4}$&$[1.40,3.27]\times10^{-4}$\\\hline
$\mathcal{B}(B^+_u\to
\rho^0\ell'^+\nu_{\ell'})$&$(1.28\pm0.18)\times10^{-4}$&$[1.49,4.32]\times10^{-4}$&$[1.48,4.45]\times10^{-4}$\\\hline
$\mathcal{B}(B^0_d\to
\rho^-\ell'^+\nu_{\ell'})$&$(2.2\pm0.4)\times10^{-4}$&$[2.78,8.02]\times10^{-4}$&$[2.77,8.32]\times10^{-4}$\\\hline\hline
\end{tabular}
\end{center}
\end{table}

For the exclusive $b\to u \ell' \nu_{\ell'}$ decays, several
branching ratios have been accurately measured by
\textit{B{\footnotesize A}B{\footnotesize AR}}, Belle and CLEO
\cite{BelleBMlnu,Babarpimlnu1,Babarpimlnu2,BabarM0mlnu,CLEOrholnu,CLEOMmlnu}.
Their averaged values from PDG \cite{PDG2006} and corresponding SM
prediction values are given in Table VII. The experimental results
are roughly consistent with the SM predictions, nevertheless there
are still windows for NP in these processes. Because many branching
ratios have been accurately measured, in order to easily obtain the
solution of the RPV coupling products, we will use the experimental
data given in Table VII, which are varied randomly within $2\sigma$
range to constrain the RPV coupling products.

\begin{figure}[t]
\begin{center}
\includegraphics[scale=0.54]{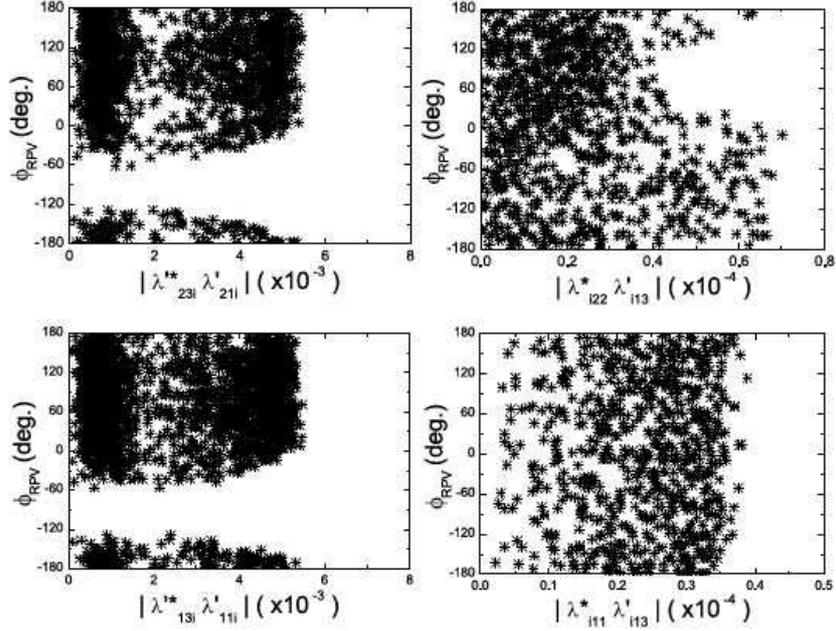}
\end{center}\vspace{-1cm}
\caption{The allowed parameter spaces for the relevant
 RPV coupling products constrained by the measurements of the exclusive $\overline{b}\to
\overline{u} \ell'^+ \nu_{\ell'}$ decays listed in Table
VII.}\label{figmuebounds}
\end{figure}

Four RPV coupling products
$\lambda'^*_{23i}\lambda'_{21i},\lambda^*_{i22}\lambda'_{i13}$ for
$\ell'=\mu$ and
$\lambda'^*_{13i}\lambda'_{11i},\lambda^*_{i11}\lambda'_{i13}$ for
$\ell'=e$ are related to fourteen exclusive $b\to u \ell'^+
\nu_{\ell'}$ decay modes. We use
$\mathcal{B}(B^+_u\to\ell'^+\nu_{\ell'})$, $\mathcal{B}(B^0_d\to
\pi^-(\rho^-)\ell'^+\nu_{\ell'})$, $\mathcal{B}(B^+_u\to
\pi^+(\rho^+)\ell'^+\nu_{\ell'})$ and their experimental data listed
in Table VII to restrict the relevant RPV parameter spaces. The
random variation of the parameters subjected to the constraints as
discussed above leads to the scatter plots displayed in Fig.
\ref{figmuebounds}.
In Fig. \ref{figmuebounds}, the RPV weak phases of the slepton
exchange couplings $\lambda^*_{i22}\lambda'_{i13}$ and
$\lambda^*_{i11}\lambda'_{i13}$ have the entirely allowed ranges
$[-180^{\circ},180^{\circ}]$, but for every RPV weak phase of the
squark exchange couplings $\lambda'^*_{23i}\lambda'_{21i}$ and
$\lambda'^*_{13i}\lambda'_{11i}$,
 there are two possible bands. For $\lambda'^*_{23i}\lambda'_{21i}$,
one band of its phase is
$\phi_{\spur{R_p}}\in[-180^{\circ},-129^{\circ}]$, another is
$\phi_{\spur{R_p}}\in[-61^{\circ},180^{\circ}]$.  And for
$\lambda'^*_{13i}\lambda'_{11i}$,  one band  is
$\phi_{\spur{R_p}}\in[-180^{\circ},-129^{\circ}]$, another is
$\phi_{\spur{R_p}}\in[-56^{\circ},180^{\circ}]$.   The magnitudes of
the squark and slepton exchange couplings have been upper limited.
The upper limits are summarized in Table VIII.
Compared with the existing bounds \cite{0406029,0406039,Allanach99},
which are estimated from the products of the smallest values of two single couplings, we get quite
strong quadric bounds on $|\lambda^*_{i22}\lambda'_{i13}|$ and
$|\lambda^*_{i11}\lambda'_{i13}|$, due to the slepton exchange
couplings.

\begin{table}[t]
\centerline{\parbox{10.8cm}{Table VIII: \small {Bounds for  the
relevant RPV coupling products by the exclusive $\overline{b}\to
\overline{u} \ell'^+ \nu_{\ell'}$ decays for 100 ${\rm GeV}$
sfermions, and previous bounds are listed for comparison
\cite{0406029,0406039,Allanach99}. }}} \vspace{0.5cm}
\begin{center}
\begin{tabular}{c|l|l}\hline\hline
Couplings&~~~Bounds~~~~~~[Processes]& Previous bounds \\
\hline $|\lambda'^*_{23i}\lambda'_{21i}|$&$\leq5.44\times
10^{-3}~[^{B^+_u\to\mu^+ \nu_\mu}_{B\to M' \mu^+\nu_\mu}]$
&$\leq2.64\times 10^{-3}$\\
$|\lambda^*_{i22}\lambda'_{i13}|$&$\leq7.00\times
10^{-5}~[^{B^+_u\to\mu^+ \nu_\mu}_{B\to M' \mu^+\nu_\mu}]$
&$\leq3.24\times 10^{-3}$\\\hline
$|\lambda'^*_{13i}\lambda'_{11i}|$&$\leq5.49\times 10^{-3}~[^{B^+_u\to e^+ \nu_e}_{B\to M' e^+\nu_e}]$&
$\leq5.4\times 10^{-3}$\\
$|\lambda^*_{i11}\lambda'_{i13}|$&$\leq3.88\times
10^{-5}~[^{B^+_u\to e^+ \nu_e}_{B\to M' e^+\nu_e}]$
&$^{\leq2.89\times 10^{-3}~ (i=2)}_{\leq6.82\times 10^{-3}~ (i=3)}
$\\\hline
\end{tabular}
\end{center}
\end{table}

Using the constrained parameter spaces shown in Fig.
\ref{figmuebounds}, we predict the RPV effects on other quantities
which have not been measured yet in the exclusive $\overline{b}\to
\overline{u} \ell'^+ \nu_{\ell'}$ decays. Our predictively numerical
results are summarized in Table IX.
\begin{table}[b]
\centerline{\parbox{17cm}{\small Table IX: The theoretical
predictions for CP averaged $\mathcal{B}$ and
$\Gamma_L^V/\Gamma_T^V$ of the exclusive $\overline{b}\to
\overline{u} \ell'^+ \nu_{\ell'}$ decays in the SM and the RPV MSSM.
The RPV MSSM predictions are obtained by the constrained regions of
the different RPV coupling products. The index  $g=1$ and $2$ for
$\ell'=e$ and $\mu$, respectively.}} \vspace{0.3cm}
\begin{center}\small{
\begin{tabular}{l|c|c|c}\hline\hline
&SM value & MSSM  value w/ $\lambda'^*_{g3i}
 \lambda'_{g1i}$& MSSM  value w/  $\lambda^*_{igg}
 \lambda'_{i13}$\\\hline
$\mathcal{B}(B^+_u\to
\mu^+\nu_\mu)$&$[2.69,5.30]\times10^{-7}$&$[1.55,3.64]\times10^{-7}$&$[0.03,16.98]\times10^{-7}$\\\hline
$\mathcal{B}(B^0_s\to K^-
\mu^+\nu_\mu)$&$[1.98,4.81]\times10^{-4}$&$[1.14,3.07]\times10^{-4}$&$[2.00,3.45]\times10^{-4}$\\\hline
$\mathcal{B}(B^0_s\to K^{*-}
\mu^+\nu_\mu)$&$[3.17,8.99]\times10^{-4}$&$[1.99,5.14]\times10^{-4}$&$[3.17,6.43]\times10^{-4}$\\\hline
$\frac{\Gamma_L}{\Gamma_T}(B^+_u\to \rho^0
\mu^+\nu_\mu)$&$[0.49,1.52]$&$\cdots\cdots$&$[0.54,0.66]$\\\hline
$\frac{\Gamma_L}{\Gamma_T}(B^0_d\to \rho^+
\mu^+\nu_\mu)$&$[0.49,1.52]$&$\cdots\cdots$&$[0.54,0.66]$\\\hline
$\frac{\Gamma_L}{\Gamma_T}(B^0_s\to K^{*-}
\mu^+\nu_\mu)$&$[0.68,1.70]$&$\cdots\cdots$&$[0.71,1.63]$\\\hline\hline
$\mathcal{B}(B^+_u\to
e^+\nu_e)$&$[6.26,12.37]\times10^{-12}$&$[3.49,8.60]\times10^{-12}$&$[6.26\times10^{-12},9.8\times10^{-7}]$\\\hline
$\mathcal{B}(B^0_s\to K^-
e^+\nu_e)$&$[1.99,4.78]\times10^{-4}$&$[1.15,3.07]\times10^{-4}$&$[2.01,3.43]\times10^{-4}$\\\hline
$\mathcal{B}(B^0_s\to K^{*-}
e^+\nu_e)$&$[3.19,8.96]\times10^{-4}$&$[1.89,5.22]\times10^{-4}$&$[3.29,6.41]\times10^{-4}$\\\hline
$\frac{\Gamma_L}{\Gamma_T}(B^+_u\to \rho^0
e^+\nu_e)$&$[0.48,1.53]$&$\cdots\cdots$&$[0.53,0.66]$\\\hline
$\frac{\Gamma_L}{\Gamma_T}(B^0_d\to \rho^+
e^+\nu_e)$&$[0.48,1.53]$&$\cdots\cdots$&$[0.53,0.66]$\\\hline
$\frac{\Gamma_L}{\Gamma_T}(B^0_s\to K^{*-}
e^+\nu_e)$&$[0.69,1.68]$&$\cdots\cdots$&$[0.73,1.67]$\\\hline
\end{tabular}}
\end{center}
\end{table}
Because the RPV effects on the exclusive $\overline{b}\to
\overline{u} \mu^+ \nu_\mu$ and $\overline{b}\to \overline{u} e^+
\nu_e$ are quite similar, as shown in Table IX, here we  give their remarks
altogether:

\begin{itemize}
\item[\textcircled{\scriptsize 1}]
For the squark exchange couplings $\lambda'^*_{g3i} \lambda'_{g1i}$,
their effects can decrease the upper limits and lower limits of
$\mathcal{B}(B^+_u\to \ell'^+\nu_{\ell'})$, $\mathcal{B}(B^0_s\to
K^{-} \ell'^+ \nu_{\ell'})$ and $\mathcal{B}(B^0_s\to K^{*-} \ell'^+
\nu_{\ell'})$, as well as shrink the allowed ranges of these branching
ratios. The  squark exchange effects are completely canceled in
$\frac{\Gamma_L}{\Gamma_T}(B\to V \ell'^+ \nu_{\ell'})$.

\item[\textcircled{\scriptsize 2}]
The slepton exchange couplings $\lambda^*_{igg}\lambda'_{i13}$,
which satisfy all present experimental constraints, could
significantly change the purely leptonic decay branching ratios
$\mathcal{B}(B^+_u\to \ell'^+\nu_{\ell'})$: They could enhance the
ratios to their experimental upper limits.  $\mathcal{B}(B^+_u\to
\mu^+\nu_{\mu})$ could be suppressed to $10^{-9}$ or enhanced to
order of $10^{-6}$, and $\mathcal{B}(B^+_u\to e^+\nu_{e})$ could be
enhanced 5 orders from order of $10^{-12}$ to order of $10^{-7}$.
The reason of these significant effects on $\mathcal{B}(B^+_u\to
\ell'^+\nu_{\ell'})$ is that the SM effective Hamiltonian is
proportional to $(\bar{b}\gamma_\mu(1-\gamma_5)u)
(\overline{\nu}_{\ell'}\gamma^\mu(1-\gamma_5)\ell')$, whose
contribution to $\mathcal{B}(B^+_u\to \ell'^+\nu_{\ell'})$ is
suppressed by $m^2_{\ell'}$ due to helicity suppression, while the
effective Hamiltonian of slepton exchange is proportional to
$(\bar{b}(1-\gamma_5)u)(\overline{\nu}_{\ell'}(1+\gamma_5)\ell')$,
whose contribution is not suppressed by $m^2_{\ell'}$. Therefore,
compared with the SM contribution,  the slepton exchange couplings
have great effects on $\mathcal{B}(B^+_u\to \ell'^+\nu_{\ell'})$.
The allowed ranges of $\mathcal{B}(B^0_s\to
K^-(K^{*-})\ell'^+\nu_{\ell'})$ and $\frac{\Gamma_L}{\Gamma_T}(B\to
V \ell'^+ \nu_{\ell'})$ are  shrunken by
$\lambda^*_{igg}\lambda'_{i13}$ couplings.
\end{itemize}

\begin{figure}[t]
\begin{center}
\includegraphics[scale=0.76]{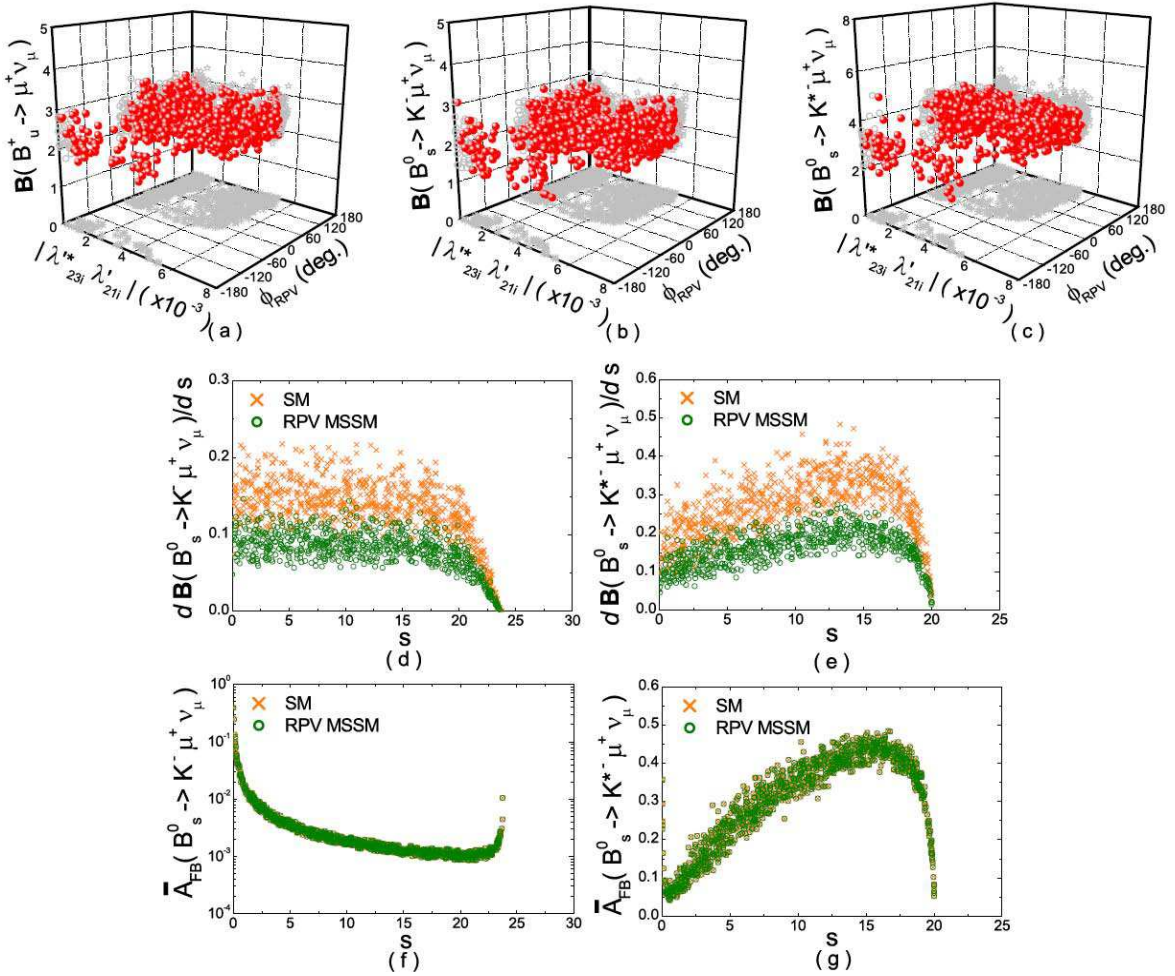}
\end{center}\vspace{-1cm}
\caption{ The effects of RPV coupling $\lambda'^*_{23i}
 \lambda'_{21i}$ on the exclusive $\bar{b}\to \bar{u} \mu^+
 \nu_\mu$ decays.
 $\mathcal{B}$ and $d\mathcal{B}/ds$ of the semileptonic decays are in unit of $10^{-4}$, and
 $\mathcal{B}(B_u^+\to \mu^+\nu_\mu)$ is in unit of $10^{-7}$.}\label{figlpslpmu}
\end{figure}

\begin{figure}[t]
\begin{center}
\includegraphics[scale=0.72]{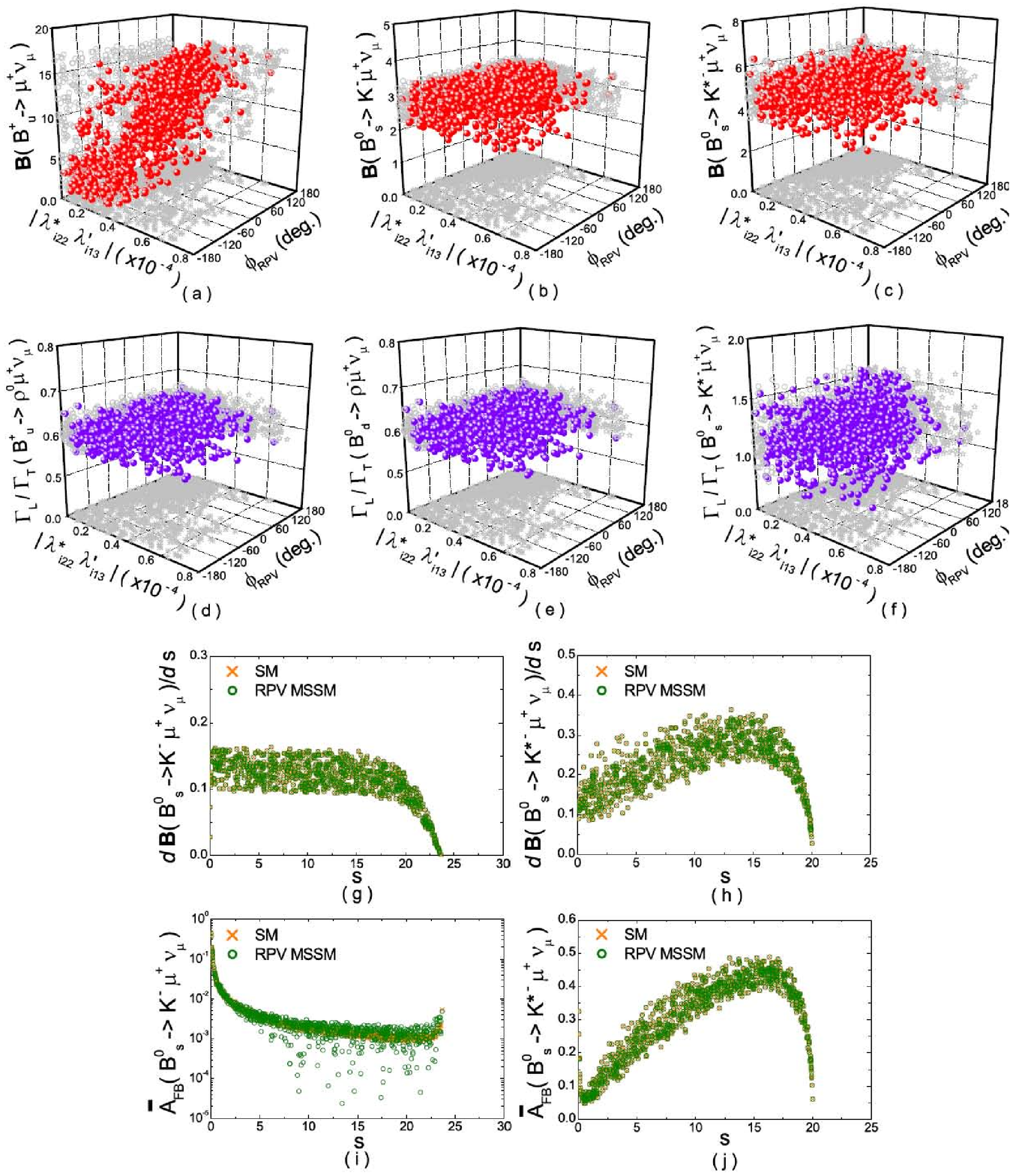}
\end{center}\vspace{-1cm}
\caption{The effects of RPV coupling $\lambda^*_{i22}
 \lambda'_{i13}$ on the exclusive $\bar{b}\to \bar{u} \mu^+
 \nu_\mu$ decays.
 $\mathcal{B}$ and $d\mathcal{B}/ds$ of the semileptonic decays are in unit of $10^{-4}$, and
 $\mathcal{B}(B_u^+\to \mu^+\nu_\mu)$ is in unit of $10^{-7}$.}\label{figlslpmu}
\end{figure}

Figs. \ref{figlpslpmu}-\ref{figlslpmu} show the RPV contributions in the
$\overline{b}\to \overline{u} \mu^+  \nu_\mu$ decays.
We view that the trends in the
changes of the physical observables with the modulus and weak phase
$\phi_{\spur{R_p}}$ of the RPV couplings by the three-dimensional
scatter plots, and we also compare the SM predictions with the RPV
MSSM predictions in $d\mathcal{B}/ds$ and
$\overline{\mathcal{A}}_{FB}$  by the two-dimensional scatter plots.
Fig. \ref{figlpslpmu} displays the $\lambda'^*_{23i}\lambda'_{21i}$
effects due to the squark exchange couplings on the exclusive
$\bar{b}\to \bar{u} \mu^+
 \nu_{\mu}$ decays.
{}From Fig. \ref{figlpslpmu}(d-e), we find the contributions of
$\lambda'^*_{23i}\lambda'_{21i}$ can suppress $d\mathcal{B}(B^0_s\to
K^- \mu^+\nu_{\mu})/ds$ and $d\mathcal{B}(B^0_s\to K^{*-}
\mu^+\nu_{\mu})/ds$, so their
   contributions are easily distinguishable from the SM predictions with
theoretical uncertainties included. However,  these contributions to  other
  observables are small, and we cannot find visible effects on $\mathcal{B}(B^+_u\to
\mu^+\nu_{\mu})$, $\mathcal{B}(B^0_s\to K^-\mu^+\nu_{\mu})$,
$\mathcal{B}(B^0_s\to K^{*-}\mu^+\nu_{\mu})$,
$\overline{\mathcal{A}}_{FB}(B^0_s\to K^-\mu^+\nu_{\mu})$ and
$\overline{\mathcal{A}}_{FB}(B^0_s\to K^{*-}\mu^+\nu_{\mu})$.
Fig. \ref{figlslpmu} presents the $\lambda^*_{i22}\lambda'_{i13}$
effects due to the slepton exchange couplings on the exclusive
$\bar{b}\to \bar{u} \mu^+
 \nu_{\mu}$ decays.  The three-dimensional scatter plot Fig. \ref{figlslpmu}(a)  shows
$\mathcal{B}(B^+_u\to\mu^+\nu_{\mu})$ correlated with
$|\lambda^*_{i22}\lambda'_{i13}|$  and its phase
$\phi_{\spur{R_p}}$, so we can see that
$\mathcal{B}(B^+_u\to\mu^+\nu_{\mu})$ is greatly increased with
$|\lambda^*_{i22}\lambda'_{i13}|$, but is insensitive to
$\phi_{\spur{R_p}}$. {}From Fig. \ref{figlslpmu}(i), we find
$\lambda^*_{i22}\lambda'_{i13}$ coupling contributions to
$\overline{\mathcal{A}}_{FB}(B^0_s\to K^-\mu^+\nu_{\mu})$ are possibly large.
There are not obvious $\lambda^*_{i22}\lambda'_{i13}$ coupling
effects, overlapping with the SM results
in $\mathcal{B}(B^0_s\to K^-\mu^+\nu_{\mu})$,
$\mathcal{B}(B^0_s\to K^{*-}\mu^+\nu_{\mu})$,
$\frac{\Gamma_L}{\Gamma_T}(B\to V\mu^+\nu_{\mu})$,
$d\mathcal{B}(B^0_s\to K^-\mu^+\nu_{\mu})/ds$,
$d\mathcal{B}(B^0_s\to K^{*-}\mu^+\nu_{\mu})/ds$ and
$\overline{\mathcal{A}}_{FB}(B^0_s\to K^{*-}\mu^+\nu_{\mu})$.

For the exclusive $\overline{b}\to \overline{u} e^+ \nu_e$  decays,
the effects of $\lambda^*_{i11}\lambda'_{i13}$ on
$\overline{\mathcal{A}}_{FB}(B^0_s\to K^-e^+\nu_{e})$ can be
distinguishible from the SM prediction, but both the SM prediction and
the RPV MSSM prediction are too small to be accessible at LHC.

\section{Summary}

In this paper, we have studied the 21 decay channels
$B^+_u\to\ell^+\nu_\ell$, $B^+_u\to\pi^0\ell^+\nu_\ell$,
 $B^0_d\to\pi^-\ell^+\nu_\ell$, $B^0_s\to K^- \ell^+\nu_\ell$,  $B^+_u\to\rho^0\ell^+\nu_\ell$,
 $B^0_d\to\rho^-\ell^+\nu_\ell$ and $B^0_s\to K^{*-} \ell^+\nu_\ell$ $(\ell=\tau,\mu,e)$ in
 the MSSM with and without $R_p$ violation.
Considering the theoretical uncertainties and the experimental
errors, we have obtained fairly constrained parameter spaces of new
physics coupling constants from the present experimental data.
Furthermore, we have predicted the charged Higgs effects and the RPV
effects on the branching ratios, the normalized FB asymmetries of
charged leptons and the ratios of longitudinal to transverse
polarization of the vector mesons, which have not been measured or
have not been well measured yet.

We have found that both the charged Higgs coupling and the slepton
exchange coupling $\lambda^*_{i33}
 \lambda'_{i13}$ have significant effects on
 $\overline{\mathcal{A}}_{FB}(B\to P\tau^+\nu_\tau)$,
and the sign of $\overline{\mathcal{A}}_{FB}(B\to P\tau^+\nu_\tau)$
could be changed by these effects. The charged Higgs effects and the
slepton exchange coupling effects are distinguishable in the purely
leptonic $B^+_u\to \mu^+\nu_\mu,e^+\nu_e$ decays.  The charged Higgs
coupling has negligible effects on
$\mathcal{B}(B^+_u\to\mu^+\nu_\mu)$ and $\mathcal{B}(B^+_u\to
e^+\nu_e)$, but the slepton exchange contributions of the RPV MSSM
are very sensitive  to $\mathcal{B}(B^+_u\to\mu^+\nu_\mu)$ and
$\mathcal{B}(B^+_u\to e^+\nu_e)$. If the enhancement of branching
ratios is not discovered in $B^+\to \mu^+\nu_\mu,e^+\nu_e$ decays,
the new limits from future experiments would constrain the slepton
exchange couplings. Otherwise, it would imply that RPV effects is
likely to be seen. We have also compared the SM predictions with the
RPV predictions of  $d\mathcal{B}/ds$ and
  $\overline{\mathcal{A}}_{FB}$  in $B \to P(V) \ell^+\nu_\ell$ decays. We have found that
the RPV couplings due to squark exchange
 are in principle distinguishable from the SM contributions at all kinematic regions
 in all eighteen semileptonic $d\mathcal{B}/ds$.
The results in this paper could be useful for probing the charged
Higgs effects and the RPV MSSM effects,  and will correlate strongly
with searches for the direct SUSY signals at future experiments, for
example, LHC and Super-$B$ Factories.

\section*{Acknowledgments}
We would like to thank Dr. R. Zwicky for useful discussions on $B_s\to
K$ form factors. The work of C.S.K. was supported in part by
CHEP-SRC and in part by the KRF Grant funded by the Korean
Government (MOEHRD) No. KRF-2005-070-C00030.  The work of Ru-Min
Wang was supported by the KRF Grant funded by the Korean Government
(MOEHRD) No. KRF-2005-070-C00030.

\end{document}